\newif\ifdraft\draftfalse 
\newif\ifcamera\cameratrue 
\newcommand\texroot{.}

\PassOptionsToPackage{final}{microtype} 
\documentclass[acmsmall,\ifcamera final\else anonymous,\ifdraft draft\else review\fi\fi]{acmart}
\usepackage{local}
\allowdisplaybreaks[3]

\pretocmd{\proof}{%
  \setlength\parindent{0pt}%
  \setlength\parskip{1em}%
}{}{\typeout{Patching proof failed.}}

\setcopyright{none}
\renewcommand\footnotetextcopyrightpermission[1]{}
\makeatletter
\fancypagestyle{firstpagestyle}{\fancyhf{}\fancyhead[L]{\ACM@linecountL}}
\makeatother
\begin{document}

\title{Probabilistic Program Equivalence for \netkat}
\titlenote{This is a preliminary draft from \today.}
\authorsaddresses{}

\author{Steffen Smolka}
\orcid{0000-0001-8625-5490}
\affiliation{%
  \institution{Cornell University}%
  \country{USA}%
}
\email{smolka@cs.cornell.edu}

\author{Praveen Kumar}
\affiliation{%
  \institution{Cornell University}%
  \country{USA}%
}
\email{praveenk@cs.cornell.edu}

\author{Nate Foster}
\affiliation{%
  \institution{Cornell University}%
  \country{USA}%
}
\email{jnfoster@cs.cornell.edu}

\author{Justin Hsu}
\affiliation{%
  \institution{Cornell University}%
  \country{USA}%
}
\email{email@justinh.su}

\author{David Kahn}
\affiliation{%
  \institution{Cornell University}%
  \country{USA}%
}
\email{dmk254@cornell.edu}

\author{Dexter Kozen}
\affiliation{%
  \institution{Cornell University}%
  \country{USA}%
}
\email{kozen@cs.cornell.edu}

\author{Alexandra Silva}
\affiliation{%
  \institution{University College London}%
  \country{UK}%
}
\email{alexandra.silva@ucl.ac.uk}

\begin{abstract}
We tackle the problem of deciding whether two probabilistic programs are
equivalent in Probabilistic \netkat, a formal language for specifying and
reasoning about the behavior of packet-switched networks.  We show that the
problem is decidable for the history-free fragment of the language by developing
an effective decision procedure based on stochastic matrices.  The main
challenge lies in reasoning about iteration, which we address by designing an
encoding of the program semantics as a finite-state absorbing Markov
chain, whose limiting distribution can be computed exactly. In an extended case
study on a real-world data center network, we automatically verify various
quantitative properties of interest, including resilience in the presence of
failures, by analyzing the Markov chain semantics.
\end{abstract}
\maketitle
\renewcommand{\shortauthors}{S. Smolka, P. Kumar, N. Foster, J. Hsu, D. Kahn, D. Kozen, and A. Silva}

\section{Introduction}
\label{sec:intro}


Program equivalence is one of the most fundamental problems in
Computer Science: given a pair of programs, do they describe the same
computation? The problem is undecidable in general, but it
can often be solved for domain-specific languages based
on restricted computational models. For example, a classical approach
for deciding whether a pair of regular expressions denote the same
language is to first convert the expressions to deterministic finite
automata, which can then be checked for equivalence in almost linear
time \cite{Tarjan75}.
In addition to the theoretical motivation, there are also many
practical benefits to studying program equivalence. Being able to
decide equivalence enables more sophisticated applications, for
instance in verified compilation and program synthesis. Less
obviously---but arguably more importantly---deciding equivalence
typically involves finding some sort of finite, explicit
representation of the program semantics. This compact encoding can
open the door to reasoning techniques and decision procedures for
properties that extend far beyond straightforward program equivalence.


With this motivation in mind, this paper tackles the problem of deciding equivalence in
Probabilistic \netkat (\probnetkat), a language for modeling and
reasoning about the behavior of packet-switched networks. As its name
suggests, \probnetkat is based on
\netkat~\cite{AFGJKSW13a,FKMST15a,compilekat}, which is in turn based
on Kleene algebra with tests (\kat), an algebraic system
combining Boolean predicates and regular expressions. \probnetkat
extends \netkat with a random choice operator and a semantics based on
Markov kernels~\cite{probnetkat-scott}. The framework can be used to
encode and reason about randomized protocols (e.g., a
routing scheme that uses random forwarding paths to balance
load~\cite{valiant82}); describe uncertainty about traffic demands (e.g., the
diurnal/nocturnal fluctuation in access patterns commonly seen in
networks for large content providers~\cite{roy15}); and model failures
(e.g., switches or links that are known to fail with some
probability~\cite{gill11}).

However, the semantics of \probnetkat is surprisingly subtle. Using
the iteration operator (i.e., the Kleene star from regular expressions), it is
possible to write programs that generate continuous distributions over
an uncountable space of packet history sets~\cite[Theorem 3]{probnetkat-cantor}.
This makes reasoning about convergence non-trivial, and requires
representing infinitary objects compactly in an implementation. To
address these issues, prior work~\cite{probnetkat-scott} developed a
domain-theoretic characterization of \probnetkat that provides notions
of approximation and continuity, which can be used to reason about
programs using only discrete distributions with finite
support. However, that work left the decidability of program
equivalence as an open problem. In this paper, we settle this question
positively for the \emph{history-free} fragment of the language, where programs
manipulate sets of packets instead of sets of packet histories (finite sequences
of packets).



Our decision procedure works by deriving a canonical, explicit representation of
the program semantics, for which checking equivalence is
straightforward. Specifically, we define a \emph{big-step} semantics
that interprets each program as a finite stochastic
matrix---equivalently, a Markov chain that transitions from input to
output in a single step. Equivalence is trivially decidable on this
representation, but the challenge lies in computing the big-step
matrix for iteration---intuitively, the finite matrix needs to somehow
capture the result of an infinite stochastic process. We address this
by embedding the system in a more refined Markov chain with a larger
state space, modeling iteration in the style of a \emph{small-step}
semantics. With some care, this chain can be transformed to an
absorbing Markov chain, from which we derive a closed form analytic
solution representing the limit of the iteration by applying
elementary matrix calculations. We prove the soundness of this
approach formally.


Although the history-free fragment of \probnetkat is a restriction of
the general language, it captures the input-output behavior of a
network---mapping initial packet states to final packet states---and
is still expressive enough to handle a wide range of problems of
interest. Many other contemporary network verification tools,
including Anteater~\cite{anteater}, Header Space Analysis~\cite{hsa},
and Veriflow~\cite{veriflow}, are also based on a history-free
model. To handle properties that involve paths (e.g., waypointing),
these tools generate a series of smaller properties to check, one for
each hop in the path. In the \probnetkat implementation, working with
history-free programs can reduce the space requirements by an
exponential factor---a significant benefit when analyzing complex
randomized protocols in large networks.


We have built a prototype implementation of our approach in OCaml. The
workhorse of the decision procedure computes a finite stochastic
matrix---representing a finite Markov chain---given an input
program. It leverages the spare linear solver UMFPACK~\cite{UMFPACK}
as a back-end to compute limiting distributions, and incorporates a
number of optimizations and symbolic techniques to compactly represent
large but sparse matrices. Although building a scalable implementation
would require much more engineering (and is not the primary focus of
this paper), our prototype is already able to handle programs of
moderate size. Leveraging the finite encoding of the semantics, we
have carried out several case studies in the context of data center
networks; our central case study models and verifies the resilience of
various routing schemes in the presence of link failures.

\paragraph{Contributions and outline.}
%
%
The main contribution of this paper is the development of a decision procedure
for history-free \probnetkat. We develop a new, tractable semantics in terms of
stochastic matrices in two steps, we establish the soundness of the semantics
with respect to \probnetkat's original denotational model, and we use the
compact semantics as the basis for building a prototype implementation with
which we carry out case studies.

In \Cref{sec:overview} and \Cref{sec:probnetkat} we
introduce \probnetkat using a simple example and motivate the need for
quantitative, probabilistic reasoning.

In \Cref{sec:big-step}, we present a semantics based on {\em finite
stochastic matrices} and show that it fully characterizes the behavior
of \probnetkat programs on packets (\Cref{thm:big-step-sound}). In
this {\em big-step} semantics, the matrices encode Markov chains over
the state space $2^\Pk$. A single step of the chain models the entire
execution of a program, going directly from the initial state
corresponding to the set of input packets the final state
corresponding to the set of output packets. Although this reduces
program equivalence to equality of finite matrices, we still need to
provide a way to explicitly \emph{compute} them. In particular, the
matrix that models iteration is given in terms of a limit.

In \Cref{sec:small-step} we derive a closed form for the big-step
matrix associated with $\polp\star$, giving an explicit representation
of the big-step semantics. It is important to note that this
is \emph{not} simply the calculation of the stationary distribution of
a Markov chain, as the semantics of $\polp\star$ is more
subtle. Instead, we define a \emph{small-step semantics}, a second
Markov chain with a larger state space such that one transition models
one iteration of $\polp\star$. We then show how to transform this
finer Markov chain into an absorbing Markov chain, which admits a
closed form solution for its limiting distribution. Together, the big-
and small-step semantics enable us to analytically compute a finite
representation of the program semantics. Directly checking these semantics for
equality yields an effective decision procedure for
program equivalence (\Cref{cor:equiv-decidable}). This is in contrast
with the previous semantics~\cite{probnetkat-cantor}, which merely
provided an approximation theorem for the semantics of iteration
$\polp\star$ and was not suitable for deciding equivalence.

In \Cref{sec:case}, we illustrate the practical applicability of our
approach by exploiting the representation of \probnetkat programs as
stochastic matrices to answer a number of questions of interest in
real-world networks. For example, we can reduce loop termination to
program equivalence: the fact that the while loop below terminates
with probability $1$ can be checked as follows:
\[
\while{\pnot{\match{\field}{0}}}{(\ptrue \mathop{\opr} \modify{\field}{0}})\quad \equiv \quad\modify{\field}{0}
\]
We also present real-world case studies that use the stochastic matrix
representation to answer questions about the resilience of data center
networks in the presence of link failures.

We discuss obstacles in extending our approach to the full \probnetkat language
in \Cref{sec:full}, including a novel automata model encodable in \probnetkat
for which equivalence seems challenging to decide. We survey
related work in \Cref{sec:rw} and conclude in \Cref{sec:conc}.

\section{Overview}
\label{sec:overview}

\newcommand\up[1]{\kw{up_{#1}}}
\newcommand\cond[3]{\pseq{\match{#1}{#2}}{#3}}

This section introduces the syntax and semantics of \probnetkat using a simple
example. We will also see how various properties, including program equivalence
and also program ordering and quantitative computations over the output
distribution, can be encoded in \probnetkat. Each of the analyses in this
section can be automatically carried out in our prototype implementation.

As our running example, consider the network shown in
\Cref{fig:example-network}. It connects Source and Destination hosts through a
topology with three switches.  Suppose we want to implement the following
policy: \emph{forward packets from the Source to the Destination}. We will start
by building a straightforward implementation of this policy in \probnetkat and
then verify that it correctly implements the specification embodied in the
policy using program equivalence. Next, we will refine our implementation to
improve its resilience to link failures and verify that the refinement is more
resilient. Finally, we characterize the resilience of both implementations
quantitatively.

\subsection{Deterministic Programming and Reasoning}
We will start with a simple deterministic program that forwards
packets from left to right through the topology. To a first
approximation, a \probnetkat program can be thought of as a random
function from input packets to output packets. We model packets as
records, with fields for standard headers such as the source address
(\src) and destination address (\dst) of a packet, as well as two
fields switch (\sw) and port (\pt) identifying the current location of
the packet. The precise field names and ranges turns out to be not so important
for our purposes; what is crucial is that the number of fields and the size of
their domains must be finite.

\netkat provides primitives $\modify{\field}{n}$ and
$\match{\field}{n}$ to modify and test the field $\field$ of an
incoming packet. A modification $\modify{\field}{n}$ returns the input
packet with the $\field$ field updated to $n$. A test
$\match{\field}{n}$ either returns the input packet unmodified if the
test succeeds, or returns the empty set if the test fails. There are
also primitives $\skp$ and $\drp$ that behave like a test that always
succeeds and fails, respectively. Programs $\polp, \polq$ are
assembled to larger programs by composing them in sequence
($\pseq{\polp}{\polq}$) or in parallel ($\punion{\polp}{\polq}$).
\netkat also provides the Kleene star operator $\polp\star$ from regular
expressions to iterate programs. \probnetkat extends \netkat with an
additional operator $\polp \oplus_r \polq$ that executes either
$\polp$ with probability $r$, or $\polq$ with probability $1-r$.

\paragraph{Forwarding.}
We now turn to the implementation of our forwarding policy. To route
packets from Source to Destination, all switches can simply forward
incoming packets out of port 2:
\begin{align*}
  \polp_1 \defeq \modify{\pt}{2} && \polp_2 \defeq \modify{\pt}{2} && \polp_3 \defeq \modify{\pt}{2}
\end{align*}
This is achieved by modifying the port field $(\pt)$. Then, to encode
the forwarding logic for all switches into a single program, we take
the union of their individual programs, after guarding the policy for
each switch with a test that matches packets at that switch:
\begin{align*}
\polp \defeq
  (\cond{\sw}{1}{p_1}) ~\pcomp~ (\cond{\sw}{2}{p_2}) ~\pcomp~ (\cond{sw}{3}{p_3})
\end{align*}
Note that we specify a policy for switch 3, even though it is
unreachable.

Now we would like to answer the following question: does our program
$\polp$ correctly forward packets from Source to Destination? Note
however that we cannot answer the question by inspecting $\polp$
alone, since the answer depends fundamentally on the network topology.

\begin{figure}
\centerline{\includegraphics[scale=0.1,draft=false]{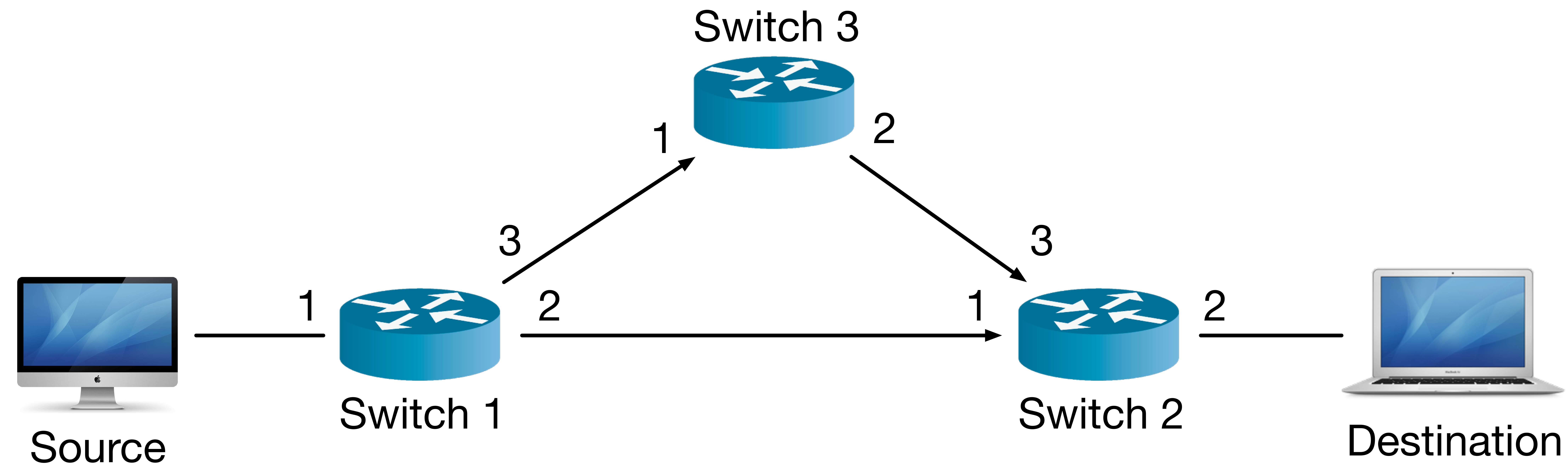}}
\vspace{-1.5ex}
\caption{Example network.}
\label{fig:example-network}
\end{figure}

\paragraph{Topology.}
Although the network topology is not programmable, we can still model
its behavior as a program. A unidirectional link matches on packets
located at the source location of the link, and updates their location
to the destination of the link. In our example network
(\Cref{fig:example-network}), the link $\ell_{ij}$ from switch $i$ to
switch $j \neq i$ is given by
\begin{align*}
\ell_{ij} &\defeq
  \pseq{\pseq{\pseq{
    \match{\sw}{i}
  }{
    \match{\pt}{j}
  }}{
    \modify{\sw}{j}
  }}{
    \modify{\pt}{i}
  }
\end{align*}
We obtain a model for the entire topology by taking the union of all its
links:
\begin{align*}
t &\defeq
  \ell_{12} \pcomp \ell_{13} \pcomp \ell_{32}
\end{align*}
Although this example uses unidirectional links,
bidirectional links can be modeled as well using a pair of
unidirectional links.

\paragraph{Network Model.}
A packet traversing the network is subject to an interleaving of
processing steps by switches and links in the network.  This is
expressible in \netkat using Kleene star as follows:
\[
  \model(p,t) \defeq \pseq{(\pseq{\polp}{t})\star}{\polp} 
\]
However, the model $\model(p,t)$ captures the behavior of the network
on arbitrary input packets, including packets that start at arbitrary
locations in the interior of the network. Typically we are interested
only in the behavior of the network for packets that originate at the
ingress of the network and arrive at the egress of the network. To
restrict the model to such packets, we can define predicates $in$ and
$out$ and pre- and post-compose the model with them:
\[
  \pseq{\pseq{in}{\model(p,t)}}{out}
\]
For our example network, we are interested in packets originating at
the Source and arriving at the Destination, so we define
\begin{align*}
  in \defeq \pseq{\match{\sw}{1}}{\match{\pt}{1}} &&
  out \defeq \pseq{\match{\sw}{2}}{\match{\pt}{2}}
\end{align*}
With a full network model in hand, we can verify that $\polp$
correctly implements the desired network policy, \ie forward packets
from Source to Destination. Our informal policy can be expressed
formally as a simple \probnetkat program:
\[
  \mathit{teleport} \defeq \pseq{\modify{\sw}{2}}{\modify{\pt}{2}}
\]
We can then settle the correctness question by checking the equivalence
\[
  \pseq{\pseq{in}{\model(p,t)}}{out} \ \equiv \ \pseq{in}{\mathit{teleport}}
\]
Previous work \cite{AFGJKSW13a,FKMST15a,compilekat} used \netkat
equivalence with similar encodings to reason about various essential
network properties including waypointing, reachability, isolation, and
loop freedom, as well as for the validation and verification of
compiler transformations. Unfortunately, the \netkat decision
procedure \cite{FKMST15a} and other state of the art network
verification tools \cite{hsa,veriflow} are fundamentally limited to
reasoning about deterministic network behaviors.

\subsection{Probabilistic Programming and Reasoning}
\label{subsec:overview-prob}
%
Routing schemes used in practice often behave
non-deterministically---e.g., they may distribute packets across
multiple paths to avoid congestion, or they may switch to backup paths
in reaction to failures. To see these sorts of behaviors in action,
let's refine our naive routing scheme $\polp$ to make it resilient to
random link failures.

\paragraph{Link Failures.}
We will assume that switches have access to a boolean flag $\up{i}$
that is true if and only if the link connected to the switch at port
$i$ is transmitting packets correctly.\footnote{Modern switches use
low-level protocols such as Bidirectional Forwarding Detection (BFD)
to maintain healthiness information about the link connected to each
port~\cite{rfc7130}.} To make the network resilient to a failure, we
can modify the program for Switch 1 as follows: if the link
$\ell_{12}$ is up, use the shortest path to Switch 2 as before;
otherwise, take a detour via Switch 3, which still forwards all
packets to Switch 2.
\[
\hat{\polp}_1 \defeq 
  \punion{(
    \pseq{\match{\up{2}}{1}}{\modify{\pt}{2}}
  )}{(
    \pseq{\match{\up{2}}{0}}{\modify{\pt}{3}}
  )}
\]
As before, we can then encode the forwarding logic for all switches into a single
program:
\begin{align*}
\hat{\polp} &\defeq
  (\cond{\sw}{1}{\hat{p_1}}) ~\pcomp~ (\cond{\sw}{2}{p_2}) ~\pcomp~ (\cond{\sw}{3}{p_3})
\end{align*}

Next, we update our link and topology encodings. A link behaves as before when
it is up, but drops all incoming packets otherwise:
\begin{align*}
\hat{\ell}_{ij} &\defeq
  \cond{\up{j}}{1}{\ell_{ij}}\\%
\end{align*}
For the purposes of this example, we will consider failures of links
connected to Switch $1$ only:
\begin{align*}
\hat{t} &\defeq
  \hat{\ell}_{12} \pcomp \hat{\ell}_{13} \pcomp \ell_{32}
\end{align*}
We also need to assume some \emph{failure model}, \ie a probabilistic
model of when and how often links fail. We will consider three failure
models:
\begin{align*}
f_0 &\defeq \pseq{
    \modify{\up{2}}{1}
  }{
    \modify{\up{3}}{1}
  }\\
f_1 &\defeq
  \bigchoice{
    f_0                                                     \withp \frac{1}{2},~
    (\modify{\up{2}}{0}) \pcomp (\modify{\up{3}}{1})        \withp \frac{1}{4},~
    (\modify{\up{2}}{1}) \pcomp (\modify{\up{3}}{0})        \withp \frac{1}{4}
  }\\
f_2 &\defeq \pseq{
    (\modify{\up{2}}{1} \oplus_{0.8} \modify{\up{2}}{0})
  }{
    (\modify{\up{2}}{1} \oplus_{0.8} \modify{\up{2}}{0})
  }
\end{align*}
Intuitively, in model $f_0$, links never fail; in $f_1$, the links
$\ell_{12}$ and $\ell_{13}$ can fail with probability $25\%$ each, but
at most one fails; in $f_2$, the links can fail independently with
probability $20\%$ each.

Finally, we can assemble the encodings of policy, topology, and
failures into a refined model:
\begin{align*}
\hat\model(p,t,f) \defeq
  ~&\kw{var}~\modify{\up{2}}{1}~\kw{in}\\
  &\kw{var}~\modify{\up{3}}{1}~\kw{in}\\
  &\model((\pseq{f}{p}), t)
\end{align*}
The refined model $\hat{\model}$ wraps our previous model $\model$
with declarations of the two local variables $\up{2}$ and $\up{3}$,
and it executes the failure model at each hop prior to switch and
topology processing. As a quick sanity check, we can verify that the
old model and the new model are equivalent in the absence of
failures, \ie under failure model
$f_0$:
\[
\model(\polp,t) \ \equiv \ \hat{\model}(\polp, \hat{t}, f_0)
\]

Now let us analyze our resilient routing scheme $\hat{\polp}$. First,
we can verify that it correctly routes packets to the Destination in the
absence of failures by checking the following equivalence:
\[
  \pseq{\pseq{in}{\hat{\model}(\hat{\polp}, \hat{t}, f_0)}}{out} \ \equiv \
  \pseq{in}{\mathit{teleport}}
\]
In fact, the scheme $\hat{\polp}$ is 1-resilient: it delivers all
packets as long as no more than $1$ link fails. In particular, it
behaves like $\mathit{teleport}$ under failure model $f_1$. In
contrast, this is not true for our naive routing scheme $\polp$:
\[
  \pseq{\pseq{in}{\hat{\model}(\hat{\polp}, \hat{t}, f_1)}}{out}
  \ \equiv \
  \pseq{in}{\mathit{teleport}}
  \ \nequiv \
  \pseq{\pseq{in}{\hat{\model}(\polp, \hat{t}, f_1)}}{out}
\]
Under failure model $f_2$, neither of the routing schemes is fully
resilient and equivalent to teleportation. However, it is reassuring
to verify that the refined routing scheme $\hat{\polp}$ performs
strictly better than the naive scheme $\polp$,
\[
  \hat\model(p,\hat{t},f_2) \ < \ \hat\model(\hat{p},\hat{t},f_2)
\]
where $p < q$ means that $q$ delivers packets with higher probability
than $p$.


Reasoning using program equivalences and inequivalences is helpful to
establish qualitative properties such as reachability properties and
program invariants.  But we can also go a step further, and compute
quantitative properties of the packet distribution generated by
a \probnetkat program. For example, we may ask for the probability
that the schemes deliver a packet originating at Source to Destination
under failure model $f_2$. The answer is $80\%$ for the naive scheme,
and $96\%$ for the resilient scheme. Such a computation might be used
by an Internet Service Provider (ISP) to check that it can meet its
service-level agreements (SLA) with customers.

In \Cref{sec:case} we will analyze a more sophisticated resilient
routing scheme and see more complex examples of qualitative and
quantitative reasoning with \probnetkat drawn from real-world data
center networks.  But first, we turn to developing the theoretical
foundations (\Cref{sec:probnetkat,sec:big-step,sec:small-step}).

\section{Background on Probabilistic NetKAT}
\label{sec:probnetkat}

In this section, we review the syntax and semantics
of \probnetkat~\cite{probnetkat-scott,probnetkat-cantor} and basic
properties of the language, focusing on the history-free fragment.  A
synopsis appears in \cref{fig:probnetkat}.

\subsection{Syntax}
A \emph{packet} $\pk$ is a record mapping a finite set of fields
$\field_1, \field_2, \dots, \field_k$ to bounded integers $n$. Fields
include standard header fields such as source (\src) and destination
(\dst) addresses, as well as two logical fields for the switch (\sw)
and port (\pt) that record the current location of the packet in the
network. The logical fields are not present in a physical network
packet, but it is convenient to model them as if they were. We write
$\pk.\field$ to denote the value of field $\field$ of $\pi$ and
$\upd{\pk}{\field}{n}$ for the packet obtained from $\pi$ by updating
field $\field$ to $n$. We let $\Pk$ denote the (finite) set of all
packets.


\probnetkat\ expressions consist of \emph{predicates} ($\preda,\predb,\ldots$)
and \emph{programs} ($\polp,\polq,\ldots$). Primitive predicates
include \emph{tests} ($\match{\field}{n}$) and the Boolean constants
\emph{false} ($\pfalse$) and \emph{true} ($\ptrue$). Compound predicates
are formed using the usual Boolean connectives of disjunction
($\punion{\preda}{\predb}$), conjunction ($\pseq{\preda}{\predb}$),
and negation ($\pnot{\preda}$). Primitive programs
include \emph{predicates} ($\preda$) and \emph{assignments}
($\modify{\field}{n}$). Compound programs are formed using the
operators \emph{parallel composition}
($\punion{\polp}{\polq}$), \emph{sequential composition}
($\pseq{\polp}{\polq}$), \emph{iteration} ($\pstar{\polp}$),
and \emph{probabilistic choice} ($\polp \opr \polq$). The full version
of the language also provides a $\pdup$ primitives, which logs the
current state of the packet, but we omit this operator from the
history-free fragment of the language considered in this paper; we discuss
technical challenges to handling full \probnetkat in \Cref{sec:full}.


The probabilistic choice operator $\polp \opr \polq$ executes $\polp$
with probability $r$ and $\polq$ with probability $1-r$, where $r$ is
rational, $0\le r\le 1$. We often use an $n$-ary version and omit the
$r$'s as in $\polp_1\oplus\cdots\oplus\polp_n$, which is interpreted
as executing one of the $\polp_i$ chosen with equal probability. This
can be desugared into the binary version.

Conjunction of predicates and sequential composition of programs use
the same syntax ($\pseq{\preda}{\predb}$ and $\pseq{\polp}{\polq}$,
respectively), as their semantics coincide.  The same is true for
disjunction of predicates and parallel composition of programs
($\punion{\preda}{\predb}$ and $\punion{\polp}{\polq}$,
respectively). The negation operator ($\neg$) may only be applied to
predicates.

The language as presented in \cref{fig:probnetkat} only includes core
primitives, but many other useful constructs can be derived.  In
particular, it is straightforward to encode conditionals and while
loops:
\begin{equation*}
\begin{aligned}
\ite{\preda}{\polp}{\polq}  &\defeqs
 \punion{\pseq{\preda}{\polp}}{\pseq{\pnot \preda}{\polq}} &\qquad\qquad
\while{\preda}{\polp}       &\defeqs
 \pseq{(\pseq{\preda}{\polp})\star}{\pnot \preda}
\end{aligned}
\end{equation*}
These encodings are well known from \kat~\cite{K97c}. Mutable and
immutable local variables can also be desugared into the core
calculus (although our implementation supports them directly):
\begin{align*}
  \kw{var}~\modify{\field}{n}~\kw{in}~\polp \defeqs
  \pseq{\pseq{\modify{\field}{n}}{p}}{\modify{\field}{0}}
\end{align*}
Here $\field$ is an otherwise unused field. The assignment
$\modify{\field}{0}$ ensures that the final value of $\field$ is
``erased'' after the field goes out of scope.

\begin{figure}[t!]
\begin{minipage}{.5175\textwidth}
\textbf{Syntax}
\[
\begin{array}{r@{\kern1ex}r@{~}c@{~}l@{\kern2em}l}
\textrm{Naturals} & n & ::=  & \mathrlap{0 \mid 1 \mid 2 \mid \cdots}\\
\textrm{Fields} & \field & ::=  & \mathrlap{\field_1 \mid \cdots \mid \field_k} \\
\textrm{Packets} & \Pk \ni \pk & ::= & \mathrlap{\sset{\field_1=n_1, \dots , \field_k = n_k}} \\
\textrm{Probabilities} & r & \in & \mathrlap{[0,1] \cap \Q}\\[1ex]
\textrm{Predicates} & \preds &
   ::= & \pfalse                       & \textit{False} \\
    & & \mid & \ptrue                  & \textit{True} \\
    & & \mid & \match{\field}{n}       & \textit{Test} \\
    & & \mid & \punion{\preda}{\predb} & \textit{Disjunction} \\
    & & \mid & \pseq{\preda}{\predb}   & \textit{Conjunction} \\
    & & \mid & \pnot{\preda}           & \textit{Negation} \\[1ex]
\textrm{Programs} & \pols &
  ::= & \preda                         & \textit{Filter} \\
    & & \mid & \modify{\field}{n}      & \textit{Assignment} \\
    & & \mid & \punion{\polp}{\polq}   & \textit{Union} \\
    & & \mid & \pseq{\polp}{\polq}     & \textit{Sequence} \\
    & & \mid & \polp \opr \polq        & \textit{Choice} \\
    & & \mid & \pstar{\polp}           & \textit{Iteration}
\end{array}\]
\end{minipage}\hfill\vrule\hfill\hspace{.5ex}\begin{minipage}{.475\textwidth}
\textbf{Semantics}\quad\fbox{\(\den{\polp} \in \pset{\Pk} \to \Dist(\pset{\Pk})\)}
\[
\def\arraystretch{1.25}
\begin{array}{r@{~~}c@{~~}l}
\den{\pfalse}(a) & \defeq &
  \delta({\emptyset})\\
\den{\ptrue}(a) & \defeq &
  \delta({a})\\
\den{\match{\field}{n}}(a) & \defeq &
  \delta({\set{\pk \in a}{\pk.f = n}}) \\
\den{\modify{\field}{n}}(a) & \defeq &
  \delta({\set{\upd{\pk}{\field}{n}}{\pk \in a }}) \\
\den{\pnot{\preda}}(a) & \defeq &
  \Dist (\lambda b. a-b)(\den{\preda}(a))\\
\den{\punion{\polp}{\polq}}(a) & \defeq &
  \Dist(\cup)(\den{\polp}(a) \times \den{\polq}(a))\\
\den{\pseq{\polp}{\polq}}(a) & \defeq &
  \lift{\den{\polq}}(\den{\polp}(a))\\
\den{\polp \mathop{\opr} \polq}(a) & \defeq &
  r \cdot \den{\polp}(a) + (1-r) \cdot \den{\polq}(a)\\
\den{\pstar\polp}(a) & \defeq & \displaystyle\bigsqcup_{n \in \N} \den{p^{(n)}}(a)\\
\multicolumn{3}{l}{\text{where } ~ p^{(0)} \defeq \ptrue, \hspace{1.5ex}
  p^{(n+1)} \defeq \punion{\ptrue}{\pseq{p}{p^{(n)}}}}
\end{array}
\]
\hrule\vspace{1ex}
\textbf{(Discrete) Probability Monad} \(\Dist\)
\[
\def\arraystretch{1.25}
\begin{array}{@{}r@{\quad}l@{}}
\text{Unit} & \delta : X \to \Dist(X)\quad\delta(x) \defeq \dirac x\\
\text{Bind} & \lift{-} : (X \to \Dist(Y)) \to \Dist(X) \to \Dist(Y)\\
& \lift{f}(\mu)(A) \defeq \sum_{x \in X} f(x)(A) \cdot \mu(x)
\end{array}
\]
\end{minipage}
\caption{\probnetkat core language: syntax and semantics.}
\label{fig:probnetkat}
\end{figure}

\subsection{Semantics}

In the full version of \probnetkat, the space $\pH$ of sets of packet
histories\footnote{A history is a non-empty finite sequence of packets
modeling the trajectory of a single packet through the network.}  is
uncountable, and programs can generate continuous distributions on
this space. This requires measure theory and Lebesgue integration for
a suitable semantic treatment. However, as programs in our
history-free fragment can generate only finite discrete distributions,
we are able to give a considerably simplified presentation (\Cref{fig:probnetkat}).
Nevertheless, the resulting semantics is a direct restriction of the
general semantics originally presented
in~\cite{probnetkat-scott,probnetkat-cantor}.

\begin{proposition}
\label{prop:old-and-new-sem}
Let $\oldden{-}$ denote the semantics defined
in \cite{probnetkat-scott}.  Then for all $\pdup$-free programs
$\polp$ and inputs $a \in \pPk$, we have
\(
  \den{\polp}(a) = \oldden{\polp}(a)
\), 
where we identify packets and histories of length one.
\end{proposition}
\begin{proof*}
The proof is given in \Cref{app:omitted-proofs}.
\end{proof*}

For the purposes of this paper, we work in the discrete space
$\pPk$, \ie, the set of sets of packets. An \emph{outcome} (denoted by
lowercase variables $a,b,c,\dots$) is a set of packets and
an \emph{event} (denoted by uppercase variables $A,B,C,\dots$) is a
set of outcomes. Given a discrete probability measure on this space,
the probability of an event is the sum of the probabilities of its
outcomes.

Programs are interpreted as \emph{Markov kernels} on the space $\pPk$.
A Markov kernel is a function $\pPk \to \Dist(\pPk)$ in the
probability (or Giry) monad $\Dist$ \cite{giry1982categorical,K81c}.
Thus, a program $\polp$ maps an input set of packets $a \in \pPk$ to a
\emph{distribution} $\den{\polp}(a) \in \Dist(\pPk)$ over output sets of packets.
The semantics uses the following probabilistic
primitives:\footnote{The same primitives can be defined for
uncountable spaces, as would be required to handle the full language.}
\begin{itemize}
  \item For a discrete measurable space $X$, $\Dist(X)$ denotes the
  set of probability measures over $X$; that is, the set of countably
  additive functions $\mu : 2^X \to [0,1]$ with $\mu(X)=1$.
  \item For a measurable function $f : X \to Y$, $\Dist(f)
  : \Dist(X) \to \Dist(Y)$ denotes the \emph{pushforward} along $f$;
  that is, the function that maps a measure $\mu$ on $X$ to
  \[
  \Dist(f)(\mu) \defeq \mu \circ f^{-1} = \lambda A \in \Sigma_Y.\ \mu(\set{x \in X}{f(x) \in A})
  \] 
  which is called the \emph{pushforward measure} on $Y$.
  \item The \emph{unit} $\delta : X \to \Dist(X)$ of the monad maps a
  point $x \in X$ to the point mass (or \emph{Dirac} measure) $\dirac
  x \in \Dist(X)$.  The Dirac measure is given by
  \[
  \dirac x (A) \defeq \ind{x \in A}
  \]
  That is, the Dirac measure is $1$ if $x \in A$ and $0$ otherwise.
  \item The \emph{bind} operation of the monad,
  \[
  \bind - : (X \to \Dist(Y)) \to \Dist(X) \to \Dist(Y)
  \]
  lifts a function $f: X \to \Dist(Y)$ with deterministic inputs to a
  function $\bind f: \Dist(X) \to \Dist(Y)$ that takes random
  inputs. Intuitively, this is achieved by averaging the output of $f$
  when the inputs are randomly distributed according to
  $\mu$. Formally,
  \begin{equation*}
    \bind f(\mu)(A) \defeq 
    \sum_{x \in X} f(x)(A) \cdot \mu(x).
  \end{equation*}
  \item Given two measures $\mu \in \Dist(X)$ and $\nu \in \Dist(Y)$,  
  \[
  \mu\times\nu \in \Dist(X \times Y)
  \]
  denotes their \emph{product measure}. This is the unique measure
  satisfying:
  \[
  (\mu\times\nu)(A \times B) = \mu(A) \cdot \nu(B)
  \]
  Intuitively, it models distributions over pairs of independent
  values.
\end{itemize}

With these primitives at our disposal, we can now make our operational
intuitions precise. Formal definitions are given
in \Cref{fig:probnetkat}. A predicate $\preda$ maps (with probability
$1$) the set of input packets $a \in \pPk$ to the subset of packets
$b \subseteq a$ satisfying the predicate. In particular, the false
primitive $\pfalse$ simply drops all packets (i.e., it returns the
empty set with probability $1$) and the true primitive $\ptrue$ simply
keeps all packets (i.e., it returns the input set with probability
$1$).  The test $\match{\field}{n}$ returns the subset of input
packets whose $\field$-field contains $n$.  Negation $\pnot{\preda}$
filters out the packets returned by $\preda$.

Parallel composition $\punion{\polp}{\polq}$ executes $\polp$ and $\polq$
independently on the input set, then returns the union of their results. Note
that packet sets do \emph{not} model nondeterminism, unlike the usual situation
in Kleene algebras---rather, they model collections of packets
traversing possibly different portions of the network simultaneously.
Probabilistic choice
$\polp \opr \polq$ feeds the input to both $\polp$ and $\polq$ and
returns a convex combination of the output distributions according to
$r$. Sequential composition $\pseq{\polp}{\polq}$ can be thought of as
a two-stage probabilistic experiment: it first executes $\polp$ on the
input set to obtain a random intermediate result, then feeds that into
$\polq$ to obtain the final distribution over outputs. The outcome of
$\polq$ needs to be averaged over the distribution of intermediate
results produced by $\polp$. It may be helpful to think about summing
over the paths in a probabilistic tree diagram and multiplying the
probabilities along each path.

We say that two programs are \emph{equivalent}, denoted
$\polp \equiv \polq$, if they denote the same Markov kernel, \ie if
$\den{\polp} = \den{\polq}$.  As usual, we expect Kleene star
$\polp\star$ to satisfy the characteristic fixed point equation
$\polp\star \equiv \punion{\ptrue}{\pseq{\polp}{\polp\star}}$, which
allows it to be unrolled ad infinitum. Thus we define it as the
supremum of its finite unrollings $\polp^{(n)}$; see
\cref{fig:probnetkat}. This supremum is taken
in a CPO $(\Dist(\pPk), \sqleq)$ of distributions that is described in
more detail in \cref{sec:measurable-space}.  The partial ordering
$\sqleq$ on packet set distributions gives rise to a partial ordering
on programs: we write $\polp \leq \polq$ iff
$\den{\polp}(a) \sqleq \den{\polq}(a)$ for all inputs
$a \in \pPk$. Intuitively, $\polp \leq \polq$ iff $\polp$ produces any
particular output packet $\pi$ with probability at most that of
$\polq$ for any fixed input.

A fact that should be intuitively clear, although it is somewhat
hidden in our presentation of the denotational semantics, is that the
predicates form a Boolean algebra:
\begin{lemma}
\label{lem:preds-boolean-algebra}
Every predicate $t$ satisfies $\den{t}(a) = \dirac{a \cap b_t}$ for a
certain packet set
$b_t \subseteq \Pk$, where
\begin{itemize}
\item $b_\pfalse = \emptyset$,
\item $b_\ptrue = \Pk$,
\item $b_{\match{\field}{n}} = \set{\pk\in\Pk}{\pk.f=n}$,
\item $b_{\pnot{\preda}} = \Pk - b_{\preda}$,
\item $b_{\punions{\preda}{\predb}} = b_{\preda} \cup b_{\predb}$, and
\item $b_{\pseq{\preda}{\predb}} = b_{\preda} \cap b_{\predb}$.
\end{itemize}
\end{lemma}
\begin{proof*}
For $\pfalse$, $\ptrue$, and $\match{\field}{n}$, the claim holds trivially.
For $\pnot{\preda}$, $\punion{\preda}{\predb}$, and $\pseq{\preda}{\predb}$,
the claim follows inductively, using that
$\Dist(f)(\dirac b) = \dirac{f(b)}$,
$\dirac{b} \times \dirac{c} = \dirac{(b,c)}$, and that
$f^\dagger(\dirac{b}) = f(b)$. The first and last equations hold because
$\langle\Dist, \delta, \lift{-}\rangle$ is a monad.
\end{proof*}

\subsection{The CPO \texorpdfstring{$(\Dist(\pPk), \sqleq)$}{of distributions}}
\label{sec:measurable-space}
The space $\pPk$ with the subset order forms a CPO
$(\pPk, \subseteq)$. Following Saheb-Djahromi~\cite{Saheb-Djahromi80},
this CPO can be lifted to a CPO $(\Dist(\pPk), \sqleq)$ on
distributions over $\pPk$. Because $\pPk$ is a finite space, the
resulting ordering $\sqleq$ on distributions takes a particularly easy
form:
\[
\mu \sqleq \nu \quad \iff \quad \mu(\sset{a}{\uparrow}) \leq \nu(\sset{a}{\uparrow}) \text{ for all } a \subseteq \Pk
\]
where $\sset{a}{\uparrow} \defeq \set{b}{a \subseteq b}$ denotes
upward closure.  Intuitively, $\nu$ produces more outputs then $\mu$.
As was shown in \cite{probnetkat-scott}, \probnetkat satisfies various
monotonicity (and continuity) properties with respect to this
ordering, including
\begin{align*}
a \subseteq a' \ \implies \ \den{\polp}(a) \sqleq \den{\polp}(a')
\qquad \text{and} \qquad
n \leq m\ \implies \ \den{\polp^{(n)}}(a) \sqleq \den{\polp^{(m)}}(a).
\end{align*}
As a result, the semantics of $\polp\star$ as the supremum of its
finite unrollings $\polp^{(n)}$ is well-defined.

While the semantics of full \probnetkat requires domain theory to give
a satisfactory characterization of Kleene star, a simpler
characterization suffices for the history-free fragment:
\begin{lemma}[Pointwise Convergence]
\label{lem:ptwise-convergence}
Let $A \subseteq \pPk$. Then for all programs $\polp$ and inputs
$a \in \pPk$,
\begin{align*}
  \den{\polp\star}(a)(A) = \lim_{n\to\infty} \den{\polp^{(n)}}(a)(A).
\end{align*}
\end{lemma}
\begin{proof*}
See \Cref{app:omitted-proofs}
\end{proof*}
This lemma crucially relies on our restrictions to $\pdup$-free programs and the
space $\pPk$. With this insight, we can now move to a concrete semantics based
on Markov chains, enabling effective computation of program semantics.

\section{Big-Step Semantics}
\label{sec:big-step}


The Scott-style denotational semantics of \probnetkat
interprets programs as Markov kernels $\pPk \to \Dist(\pPk)$.
Iteration is characterized in terms of approximations in a CPO $(\Dist(\pPk), \sqleq)$ of distributions.
In this section we relate this semantics to a Markov chain semantics on a state
space consisting of finitely many packets.

Since the set of packets $\Pk$ is finite, so is its powerset $\pPk$.
Thus any distribution over packet sets is discrete and can be
characterized by a \emph{probability mass function}, \ie a function
\[
  f : \pPk \to [0,1], \qquad \sum_{b \subseteq \Pk} f(b) = 1
\]
It is convenient to view $f$ as a \emph{stochastic vector},
\ie a vector of non-negative entries that sums to $1$.
The vector is indexed by packet sets $b \subseteq \Pk$ with $b$-th
component $f(b)$. A program, being a function that maps inputs $a$ to
distributions over outputs, can then be organized as a square matrix
indexed by $\Pk$ in which the stochastic vector corresponding to input
$a$ appears as the $a$-th row.

Thus we can interpret a program $\polp$ as a matrix $\bden{\polp} \in
[0,1]^{\pPk\times\pPk}$ indexed by packet sets, where the matrix entry
$\bden{\polp}_{ab}$ denotes the probability that program $\polp$
produces output $b \in \pPk$ on input $a \in \pPk$. The rows of
$\bden{\polp}$ are stochastic vectors, each encoding the output
distribution corresponding to a particular input set $a$. Such a
matrix is called (right-)stochastic. We denote by $\mathbb{S}(\pPk)$
the set of right-stochastic square matrices indexed by $\pPk$.

\begin{figure}
\raggedright\qquad\fbox{$\bden{\polp}~\in~\mathbb{S}(\pPk)$}\\[.1em]
\begin{minipage}{.45\textwidth}
\begin{align*}
  \bden{\pfalse}_{ab} &\!\!\defeqs\!\! \ind{b = \emptyset}\\
  \bden{\ptrue}_{ab} &\!\!\defeqs\!\! \ind{a = b}\\
  \bden{\match{f}{n}}_{ab} &\!\!\defeqs\!\! \ind{b = \set{\pk \in a}{\pk.f = n}}\\
  \bden{\pnot{\preda}}_{ab} &\!\!\defeqs\!\! \ind{b \subseteq a} \cdot \bden{\preda}_{a,a-b}\\
  \bden{\modify{f}{n}}_{ab} &\!\!\defeqs\!\! \ind{b = \set{\pk[f:=n]}{\pk \in a}}
\end{align*}
\end{minipage}~\vrule~\begin{minipage}{.54\textwidth}
\begin{align*}
  \bden{\punion{\polp}{\polq}}_{ab} &\!\!\defeqs\!\!
    \sum_{c,d} \ind{c \cup d = b} \cdot \bden{\polp}_{a,c} \cdot \bden{\polq}_{a,d}\\
  \bden{\pseq{\polp}{\polq}} &\!\!\defeqs\!\! \bden{\polp} \cdot \bden{\polq}\\
  \bden{\polp \oplus_r \polq} &\!\!\defeqs\!\! r \cdot \bden{\polp} + (1-r) \cdot \bden{\polq}\\
  \bden{\polp\star}_{ab} &\!\!\defeqs\!\! \lim_{n \to \infty} \bden{p^{(n)}}_{ab}
\end{align*}
\end{minipage}
\caption{Big-Step Semantics: $\bden{\polp}_{ab}$ denotes the probability that program
$\polp$ produces output $b$ on input $a$.}
\label{fig:big-step}
\end{figure}

The interpretation of programs as stochastic matrices is largely
straightforward and given formally in \cref{fig:big-step}. At a high
level, deterministic program primitives map to simple
$(0,1)$-matrices, and program operators map to operations on
matrices. For example, the program primitive $\pfalse$ is interpreted
as the stochastic matrix
\begin{align}
\label{eq:smatric=MC}
\setlength{\kbcolsep}{0pt}
\setlength{\kbrowsep}{0pt}
\setlength{\arraycolsep}{2pt}
\def\arraystretch{1.2}
\bden{\pfalse} =
\kbordermatrix{
          & \emptyset & b_2     & \ldots & b_n     \\
\emptyset & 1         & 0       & \cdots & 0       \\
\vdots    & \vdots    & \vdots  & \ddots & \vdots  \\
a_n       & 1         & 0       & \cdots & 0
}
&&&
\begin{tikzpicture}[font=\footnotesize,baseline=-5.5ex]
\begin{scope}[execute at begin node=$, execute at end node=$, node distance=2em,
  ]
  \node (A2) {a_2};
  \node[below of=A2,draw=none] (A3) {\varvdots};
  \node[below of=A3] (AN) {a_n};
  \node[right of=A3, node distance=6em] (empty) {a_1 = \emptyset};
\end{scope}
\path[->,>=stealth,semithick]
  (A2) edge[bend left=15]                           node[above]{1} (empty)
  (AN) edge[bend right=15]                          node[below]{1} (empty)
  (empty) edge[loop right] node{1} (empty);
\end{tikzpicture}
\end{align}
that moves all probability mass to the $\emptyset$-column, and the
primitive $\ptrue$ is the identity matrix. The formal definitions are
given in \cref{fig:big-step} using Iverson brackets: $\ind{\varphi}$
is defined to be $1$ if $\varphi$ is true, or $0$ otherwise.

As suggested by the picture in \cref{eq:smatric=MC}, a stochastic
matrix $B \in \mathbb{S}(\pPk)$ can be viewed as a
\emph{Markov chain} (MC), a probabilistic transition system with state space $\pPk$
that makes a random transition between states at each time step. The matrix entry
$B_{ab}$ gives the probability that, whenever the system is in state $a$, it transitions to state $b$ in the next time step.
Under this interpretation, sequential composition becomes matrix
product: a step from $a$ to $b$ in $\bden{\pseq{\polp}{\polq}}$
decomposes into a step from $a$ to some intermediate state $c$ in
$\bden{\polp}$ and a step from $c$ to the final state $b$ in
$\bden{\polq}$ with probability
\begin{align*}
\bden{\pseq{\polp}{\polq}}_{ab} &= \sum_{c} \bden{\polp}_{ac} \cdot \bden{\polq}_{cb}
= (\bden{\polp} \cdot \bden{\polq})_{ab}.
\end{align*}

\subsection{Soundness}
The main theoretical result of this section is that the finite matrix
$\bden{\polp}$ fully characterizes the behavior of a program $\polp$
on packets.
\begin{theorem}[Soundness]
\label{thm:big-step-sound}
For any program $\polp$ and any sets $a,b \in \pPk$,
$\bden{\polp\star}$ is well-defined, $\bden{\polp}$ is a stochastic matrix,
and  $\bden{\polp}_{ab} = \den{\polp}(a)(\sset{b})$.
\end{theorem}
\begin{proof*}
It suffices to show the equality $\bden{\polp}_{ab}
= \den{\polp}(a)(\sset{b})$; the remaining claims then follow by
well-definedness of $\den{-}$. The equality is shown using
\Cref{lem:ptwise-convergence} and a routine induction on $\polp$:

For $\polp = \pfalse, \ptrue, \match{f}{n}, \modify{f}{n}$ we have
\[
\den{\polp}(a)(\sset{b}) = \dirac{c}(\sset{b}) = \ind{b = c} = \bden{\polp}_{ab}
\]
for $c = \emptyset, a, \set{\pk \in a}{\pk.f = n}, \set{\pk[f:=n]}{\pk \in a}$,
respectively.

For $\pnot{\preda}$ we have,
\[\begin{array}{rl@{\quad}l@{\quad}l}
\bden{\pnot{\preda}}_{ab}
&= \ind{b \subseteq a} \cdot \bden{\preda}_{a,a-b}\\
&= \ind{b \subseteq a} \cdot \den{\preda}(a)(\sset{a-b}) &&\text{(IH)}\\
&= \ind{b \subseteq a} \cdot \ind{a-b=a \cap b_t}        &&\text{(\Cref{lem:preds-boolean-algebra})}\\
&= \ind{b \subseteq a} \cdot \ind{a-b=a - (\Hist - b_t)}\\
&= \ind{b=a \cap (H-b_t)}\\
&= \den{\pnot{\preda}}(a)(b)                              &&\text{(\Cref{lem:preds-boolean-algebra})}
\end{array}
\]

For $\punion{\polp}{\polq}$, letting $\mu = \den{\polp}(a)$ and
$\nu = \den{\polq}(a)$ we have
\[\begin{array}{rl@{\quad}l@{\quad}l}
\den{\punion{\polp}{\polq}}(a)(\sset{b})
&= (\mu \times \nu)(\set{(b_1,b_2)}{b_1 \cup b_2 = b})\\
&= \sum_{b_1,b_2} \ind{b_1 \cup b_2 = b} \cdot (\mu \times \nu)(\sset{(b_1,b_2)})
  && \\ 
&= \sum_{b_1,b_2} \ind{b_1 \cup b_2 = b} \cdot
  \mu(\sset{b_1}) \cdot \nu(\sset{b_2})\\
&= \sum_{b_1,b_2} \ind{b_1 \cup b_2 = b} \cdot
  \bden{p}_{ab_1} \cdot \bden{q}_{ab_2} &&\text{(IH)}\\
&= \bden{\punion{\polp}{\polq}}_{ab}
\end{array}\]
where we use in the second step that $b \subseteq \Pk$ is finite, thus
$\set{(b_1,b_2)}{b_1 \cup b_2 = b}$ is finite.

For $\pseq{\polp}{\polq}$, let $\mu = \den{\polp}(a)$ and $\nu_c = \den{\polq}(c)$
and recall that $\mu$ is a discrete distribution on $\pset\Pk$. Thus
\[\begin{array}{rl@{\quad}l@{\quad}l}
\den{\pseq{\polp}{\polq}}(a)(\sset{b})
&= \sum_{c \in \pset\Pk} \nu_c(\sset{b}) \cdot \mu(\sset{c})\\
&= \sum_{c \in \pset\Pk} \bden{\polq}_{c,b} \cdot \bden{\polp}_{a,c}\\
&= \bden{\pseq{\polp}{\polq}}_{ab}.
\end{array}
\]

For $\polp \oplus_r \polq$, the claim follows directly from the induction hypotheses.

Finally, for $\polp\star$, we know that
$\bden{\polp^{(n)}}_{ab} = \den{\polp^{(n)}}(a)(\sset{b})$
by induction hypothesis. The key to proving the claim is
\Cref{lem:ptwise-convergence}, which allows us to take the limit
on both sides and deduce
\begin{equation*}
  \bden{\polp\star}_{ab} =
  \lim_{n \to \infty} \bden{\polp^{(n)}}_{ab} =
  \lim_{n \to \infty} \den{\polp^{(n)}}(a)(\sset{b}) =
  \den{\polp\star}(a)(\sset{b}). \qedhere
\end{equation*}
\end{proof*}

Together, these results reduce the problem of checking program
equivalence for $\polp$ and $\polq$ to checking equality of the
matrices produced by the big-step semantics, $\bden{\polp}$ and
$\bden{\polq}$.
\begin{corollary}
\label{cor:den-equal-iff-bden-equal}
For programs $\polp$ and $\polq$, $\den{\polp} = \den{\polq}$ if and only if
$\bden{\polp} = \bden{\polq}$.
\end{corollary}
\begin{proof*}
Follows directly from \Cref{thm:big-step-sound}.
\end{proof*}
Unfortunately, $\bden{\polp\star}$ is defined in terms of a
limit. Thus, it is not obvious how to compute the big-step matrix in
general. The next section is concerned with finding a closed form for
the limit, resulting in a representation that can be effectively
computed, as well as a decision procedure.

\section{Small-Step Semantics}
\label{sec:small-step}
This section derives a closed form for $\bden{\polp\star}$, allowing to compute
$\bden{-}$ explicitly. This yields an effective mechanism for checking
program equivalence on packets.

In the ``big-step'' semantics for \probnetkat, programs are
interpreted as Markov chains over the state space $\pPk$, such that a
single step of the chain models the entire execution of a program,
going directly from some initial state $a$ (corresponding to the set
of input packets) to the final state $b$ (corresponding to the set of
output packets).
Here we will instead take a ``small-step'' approach and design a Markov chain
such that one transition models one iteration of $\polp\star$.

\begin{figure}
\begin{tikzpicture}[->,auto,semithick]
  \tikzstyle{config}=[node distance=5em]
  \tikzstyle{edge}=[]

  \node[config]   (C1)                  {\config{\polp\star, a, b}};
  \node[config]   (C2) [right = of C1]  {\config{\ptrue\pcomp\pseq{\polp}{\polp\star}, a, b}};
  \node[config]   (C3) [right = of C2]  {\config{\pseq{\polp}{\polp\star}, a, b \cup a}};
  \node[config]   (C4) [below = of C3]  {\config{\polp\star, a', b \cup a}};

  \path (C1) edge [dashed, thick]  node [edge] {$1$}                   (C2)
        (C2) edge [dashed, thick]  node [edge] {$1$}                   (C3)
        (C3) edge [dashed, thick]  node [edge] {$\bden{\polp}_{a,a'}$} (C4)
        (C1) edge [sloped]         node [edge] {$\bden{\polp}_{a,a'}$} (C4);
\end{tikzpicture}
\caption{The small-step semantics is given by a Markov
  chain whose states are configurations of the form
  $\config{\textit{program},\textit{input set},\textit{output accumulator}}$.
  The three dashed arrows can be collapsed into the single solid arrow,
  rendering the program component superfluous.}
\label{fig:small-step}
\end{figure}
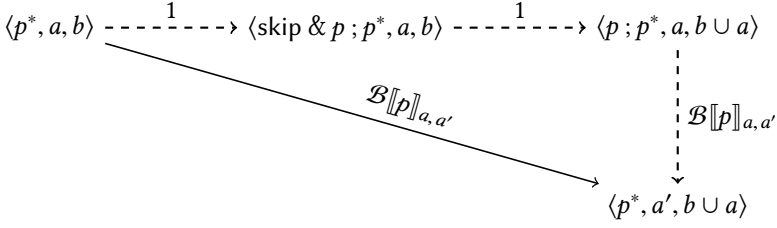

To a first approximation, the states (or configurations) of our probabilistic
transition system are triples $\config{\polp,a,b}$, consisting of the
program $\polp$ we mean to execute, the current set of (input) packets $a$,
and an accumulator set $b$ of packets output so far. The execution of
$\polp\star$ on input $a \subseteq \Pk$ starts from the initial state
$\config{\polp\star,a,\emptyset}$. It proceeds by unrolling $\polp\star$
according to the characteristic equation
$\polp\star \equiv \punion{\ptrue}{\pseq{\polp}{\polp\star}}$ with probability 1:
\begin{align*}
  \config{\polp\star,a,\emptyset} \stepsto{1}
  \config{\punion{\ptrue}{\pseq{\polp}{\polp\star}},a,\emptyset}
\end{align*}
To execute a union of programs, we must execute both programs on the input
set and take the union of their results. In the case of
$\punion{\ptrue}{\pseq{\polp}{\polp\star}}$, we can immediately execute
$\ptrue$ by outputting the input set with probability 1,
leaving the right hand side of the union:
\begin{align*}
  \config{\punion{\ptrue}{\pseq{\polp}{\polp\star}},a,\emptyset} \stepsto{1}
  \config{\pseq{\polp}{\polp\star},a,a}
\end{align*}
To execute the sequence $\pseq{\polp}{\polp\star}$, we first execute $\polp$
and then feed its (random) output into $\polp\star$:
\begin{align*}
  \forall a':\quad
  \config{\pseq{\polp}{\polp\star},a,a} \stepsto{\bden{\polp}_{a,a'}}
  \config{\polp\star,a',a}
\end{align*}
At this point the cycle closes and we are back to executing $\polp\star$, albeit
with a different input set $a'$ and some accumulated outputs.
The structure of the resulting Markov chain is shown in \cref{fig:small-step}.

At this point we notice that the first two steps of execution are deterministic,
and so we can collapse all three steps into a single one, as illustrated in
\cref{fig:small-step}. After this simplification, the program component of the
states is rendered obsolete since it remains constant across transitions. Thus
we can eliminate it, resulting in a Markov chain over the state space
$\pPk \times \pPk$. Formally, it can be defined concisely as
\begin{align*}
  \sden{\polp} &~\in~ \mathbb{S}(\pPk\times\pPk)\\
  \sden{\polp}_{(a,b),(a',b')} &\defeqs \ind{b' = b \cup a} \cdot \bden{\polp}_{a,a'}
\end{align*}
As a first sanity check, we verify that the matrix $\sden{\polp}$ defines indeed
a Markov chain:
\begin{lemma}
\label{lem:small-step-stochastic}
$\sden{\polp}$ is stochastic.
\end{lemma}
\begin{proof*} For arbitrary $a,b \subseteq \Pk$, we have
\begin{align*}
\sum_{a',b'}\sden{\polp}_{(a,b),(a',b')}
&= \sum_{a',b'} \ind{b' = a \cup b} \cdot \bden{p}_{a,a'}\\
&= \sum_{a'} \Big(\sum_{b'} \ind{b' = a \cup b} \Big) \cdot \bden{p}_{a,a'}\\
&= \sum_{a'} \bden{p}_{a,a'} = 1
\end{align*}
where, in the last step, we use that $\bden{\polp}$ is stochastic (\Cref{thm:big-step-sound}).
\end{proof*}

Next, we show that steps in $\sden{\polp}$ indeed model iterations of
$\polp\star$. Formally, the $(n+1)$-step of $\sden{\polp}$ is
equivalent to the big-step behavior of the $n$-th unrolling of
$\polp\star$ in the following sense:
\begin{proposition}
\label{prop:small-step-characterization}
$\bden{\polp^{(n)}}_{a,b} = \sum_{a'} \sden{\polp}^{n+1}_{(a,\emptyset),(a',b)}$
\end{proposition}
\begin{proof*}
Naive induction on the number of steps $n \geq 0$ fails, because the hypothesis
is too weak. We must first generalize it to apply to arbitrary start states in
$\sden{\polp}$, not only those with empty accumulator.
The appropriate generalization of the claim turns out to be:
\begin{lemma}\label{lem:star-mc-char}
  Let $p$ be program.
  Then for all $n \in \N$ and $a,b,b' \subseteq \Pk$,\[
  \sum_{a'} \ind{b' = a' \cup b} \cdot \bden{\polp^{(n)}}_{a,a'} =
  \sum_{a'} \sden{\polp}^{n+1}_{(a,b),(a',b')}
  \]
\end{lemma}
\begin{proof*}
By induction on $n \geq 0$. For $n=0$, we have
\begin{align*}
\sum_{a'} \ind{b' = a' \cup b} \cdot \bden{\polp^{(n)}}_{a,a'}
&= \sum_{a'} \ind{b' = a' \cup b} \cdot \bden{\ptrue}_{a,a'}\\
&= \sum_{a'} \ind{b' = a' \cup b} \cdot \ind{a=a'}\\
&= \ind{b' = a \cup b}\\
&= \ind{b' = a \cup b} \cdot \sum_{a'} \bden{\polp}_{a,a'}\\
&= \sum_{a'} \sden{\polp}_{(a,b),(a',b')}
\end{align*}

In the induction step ($n > 0$),
\begin{align*}
&\phantom{={}} \sum_{a'} \ind{b' = a' \cup b} \cdot
  \bden{\polp^{(n)}}_{a,a'} \\
&= \sum_{a'} \ind{b' = a' \cup b} \cdot
  \bden{\punion{\ptrue}{\pseq{p}{p^{(n-1)}}}}_{a,a'} \\
&= \sum_{a'} \ind{b' = a' \cup b} \cdot
  \sum_{c} \ind{a' = a \cup c} \cdot \bden{\pseq{p}{p^{(n-1)}}}_{a,c} \\
&= \sum_{c} \left(\sum_{a'} \ind{b' = a' \cup b} \cdot \ind{a' = a \cup c} \right)
  \cdot \sum_k \bden{p}_{a,k} \cdot \bden{p^{(n-1)}}_{k,c} \\
&= \sum_{c,k} \ind{b' = a \cup c \cup b} \cdot
  \bden{\polp}_{a,k} \cdot \bden{p^{(n-1)}}_{k,c} \\
&= \sum_{k} \bden{\polp}_{a,k} \cdot
   \sum_{a'} \ind{b' = a' \cup (a \cup b)} \cdot \bden{p^{(n-1)}}_{k,a'} \\
&= \sum_{k} \bden{\polp}_{a,k} \cdot
   \sum_{a'} \sden{\polp}^{n}_{(k,a \cup b),(a',b')} \\
&= \sum_{a'}\sum_{k_1,k_2} \ind{k_2 = a \cup b} \cdot \bden{\polp}_{a,k_1} \cdot
   \sden{\polp}^{n}_{(k_1,k_2),(a',b')} \\
&= \sum_{a'}\sum_{k_1,k_2} \sden{\polp}_{(a,b)(k_1,k_2)} \cdot
   \sden{\polp}^{n}_{(k_1,k_2),(a',b')} \\
&= \sum_{a'} \sden{\polp}^{n+1}_{(a,b),(a',b')}
\qedhere
\end{align*}
\end{proof*}
\Cref{prop:small-step-characterization} then follows by instantiating
\Cref{lem:star-mc-char} with $b=\emptyset$.
\end{proof*}

\subsection{Closed form}
\label{sec:closed-form}
Let $(a_n,b_n)$ denote the random state of the Markov chain
$\sden{\polp}$ after taking $n$ steps starting from
$(a,\emptyset)$. We are interested in the distribution of $b_n$ for
$n\to\infty$, since this is exactly the distribution of outputs
generated by $\polp\star$ on input $a$
(by \Cref{prop:small-step-characterization} and the definition of
$\bden{\polp\star}$).  Intuitively, the $\infty$-step behavior of
$\sden{\polp}$ is equivalent to the big-step behavior of $\polp\star$.
The limiting behavior of finite state Markov chains has been
well-studied in the literature (e.g., see \cite{kemeny1960finite}),
and we can exploit these results to obtain a closed form by massaging
$\sden{\polp}$ into a so called \emph{absorbing Markov chain}.

A state $s$ of a Markov chain $T$ is called \emph{absorbing} if it
transitions to itself with probability 1:
\begin{align*}
\begin{tikzpicture}[->,>=stealth,shorten >=0.3pt,semithick,baseline=-.5ex]
\tikzstyle{every state}=[minimum size=1.5em]
\tikzset{every loop/.style={in=20,out=-20,looseness=8}}
\node[state] (s) {$s$};
\path (s) edge [loop right] node {$1$} (s);
\end{tikzpicture}
&&&\text{(formally: $T_{s,s\smash{'}} = \ind{s=s'}$)}
\end{align*}
A Markov chain $T \in \mathbb{S}(S)$ is called \emph{absorbing} if each state can reach
an absorbing state:
\begin{align*}
  \forall s \in S.\ \exists s' \in S, n \geq 0.\ T^n_{s,s'} > 0 \text{ and } T_{s',s'} = 1
\end{align*}
The non-absorbing states of an absorbing MC are called \emph{transient}. Assume
$T$ is absorbing with $n_t$ transient states and $n_a$ absorbing states. After
reordering the states so that absorbing states appear before transient
states, $T$ has the form
\begin{align*}
  T = \begin{bmatrix}
    I & 0\\
    R & Q
  \end{bmatrix}
\end{align*}
where $I$ is the $n_a \times n_a$ identity matrix, $R$ is an
$n_t \times n_a$ matrix giving the probabilities of transient states transitioning
to absorbing states, and $Q$ is an $n_t \times n_t$ square matrix specifying
the probabilities of transient states transitioning to transient states. Absorbing
states never transition to transient states, thus the $n_a \times n_t$ zero
matrix in the upper right corner.

No matter the start state, a finite state absorbing MC always ends up in an
absorbing state eventually, \ie the limit $T^\infty \defeq \lim_{n\to\infty} T^n$
exists and has the form
\begin{align*}
  T^\infty = \begin{bmatrix}
    I & 0\\
    A & 0
  \end{bmatrix}
\end{align*}
for an $n_t\times n_a$ matrix $A$ of so called \emph{absorption
probabilities}, which can be given in closed form:
\begin{align*}
  A = (I + Q + Q^2 + \dots)\,R
\end{align*}
That is, to transition from a transient state to an absorbing state, the MC
can first take an arbitrary number of steps between transient states, before
taking a single and final step into an absorbing state. The infinite sum
$X \defeq \sum_{n \geq 0} Q^n$ satisfies $X = I + QX$, and solving for $X$ we
get
\begin{align}
  X = (I - Q)^{-1} \quad \text{and} \quad A = (I - Q)^{-1} R\label{eq:inverse}
\end{align}
(We refer the reader to \cite{kemeny1960finite} or Lemma \ref{lem:inverse} in Appendix \ref{app:omitted-proofs}
for the proof that the inverse must exist.)

Before we apply this theory to the small-step semantics $\sden{-}$, it will be
useful to introduce some MC-specific notation. Let $T$ be an MC. We write
$s \reaches{T}_n s'$ if $s$ can reach $s'$ in precisely $n$ steps, \ie if $T^n_{s,s'}>0$;
and we write $s \reaches{T} s'$ if $s$ can reach $s'$ in any number of steps,
\ie if $T^n_{s,s'}>0$ for any $n \geq 0$. Two states are said to communicate, denoted
$s \communicate{T} s'$, if $s \reaches{T} s'$ and $s' \reaches{T} s$. The
relation $\communicate{T}$ is an equivalence relation, and its equivalence
classes are called communication classes. A communication class is called absorbing
if it cannot reach any states outside the class. We sometimes write
$\prob{s \reaches{T}_n s'}$ to denote the probability $T^n_{s,s'}$. For the
rest of the section, we fix a program $\polp$ and abbreviate
$\bden{\polp}$ as $B$ and $\sden{\polp}$ as S.

Of central importance are what we will call the \emph{saturated} states of $S$:
\begin{definition}
\label{def:saturated-state}
A state $(a,b)$ of $S$ is called \emph{saturated} if the accumulator
$b$ has reached its final value, \ie if $(a,b) \reaches{S} (a',b')$ implies
$b'=b$.
\end{definition}
Once we have reached a saturated state, the output of $\polp\star$ is determined.
The probability of ending up in a saturated state with accumulator $b$,
starting from an initial state $(a,\emptyset)$, is
\begin{equation*}
  \lim_{n \to \infty} \sum_{a'} S^n_{(a,\emptyset),(a',b)}
\end{equation*}
and indeed this is the probability that $\polp\star$
outputs $b$ on input $a$ by \Cref{prop:small-step-characterization}.
Unfortunately, a saturated state is not necessarily absorbing. To see this,
assume there exists only a single field $\field$ ranging over $\sset{0,1}$ and
consider the program
$p\star = (\modify{\field}{0} \oplus_{1/2} \modify{\field}{1})\star $.
Then $S$ has the form
\begin{center}
\begin{tikzpicture}[->,>=stealth,shorten >=0.5pt,auto,node distance=2cm,
                    semithick]
  \tikzstyle{every state}=[draw=none,text=black]

\matrix (M) [matrix of nodes, row sep=1ex, column sep=2.5em] {
                  & $0,0$ & $0,\sset{0,1}$\\
    $0,\emptyset$ &       &               \\
                  & $1,0$ & $1,\sset{0,1}$\\
};

  \path (M-2-1) edge                                        (M-1-2.west)
        (M-2-1) edge                                        (M-3-2.west)
        (M-1-2) edge                                        (M-3-2)
        (M-1-2) edge[loop right]                            (M-1-2)
        (M-3-2) edge                                        (M-3-3)
        (M-1-3) edge[<->]                                   (M-3-3)
        (M-1-3) edge[loop right,looseness=5.5,in=-11,out=11]  (M-1-3)
        (M-3-3) edge[loop right,looseness=5.5,in=-11,out=11]  (M-3-3);

  \draw (M-3-2) edge[shorten >=-2pt] (M-1-3.south west);
\end{tikzpicture}
\end{center}
where all edges are implicitly labeled with $\frac{1}{2}$, $0$ denotes
the packet with $\field$ set to $0$ and $1$ denotes the packet with $\field$ set to $1$,
and we omit states not reachable from $(0,\emptyset)$. The two right most states
are saturated; but they communicate and are thus not absorbing.

We can fix this by defining the auxiliary matrix $U \in \mathbb{S}(\pPk \times \pPk)$ as
\begin{align*}
  U_{(a,b),(a', b')} \defeq \ind{b'=b} \cdot \begin{cases}
  \ind{a'=\emptyset}
    &\text{if $(a,b)$ is saturated}  \\
  \ind{a'=a}
    &\text{else}
  \end{cases}
\end{align*}
It sends a saturated state $(a,b)$ to the canonical saturated state $(\emptyset,b)$,
which is always absorbing; and it acts as the identity on all other states.
In our example, the modified chain $SU$ looks as follows:
\begin{center}
\begin{tikzpicture}[->,>=stealth,shorten >=0.5pt,auto,node distance=2cm,
                    semithick]
  \tikzstyle{every state}=[draw=none,text=black]

\matrix (M) [matrix of nodes, row sep=1ex, column sep=2.5em] {
                  & $0,0$ & $0,\sset{0,1}$  &                       \\
    $0,\emptyset$ &       &                 & $\emptyset,\sset{0,1}$\\
                  & $1,0$ & $1,\sset{0,1}$  &                       \\
};

  \path (M-2-1) edge                                        (M-1-2.west)
        (M-2-1) edge                                        (M-3-2.west)
        (M-1-2) edge                                        (M-3-2)
        (M-1-2) edge[loop right]                              (M-1-2)
        (M-3-2) edge                                          (M-3-3)
        (M-1-3) edge                                     (M-2-4)
        (M-3-3) edge  (M-2-4)
        (M-2-4) edge[loop right,looseness=5.5,in=-11,out=11]  (M-2-4);

  \draw (M-3-2) edge[shorten >=-2pt] (M-1-3.south west);
\end{tikzpicture}
\end{center}

To show that $SU$ is always an absorbing MC, we first observe:
\begin{lemma}
\label{lem:monotone-chains}
  $S$, $U$, and $SU$ are \emph{monotone} in the following sense:
  $(a,b) \reaches{S} (a',b')$ implies $b \subseteq b'$ (and similarly
  for $U$ and $SU$).
\end{lemma}
\begin{proof*}
  For $S$ and $U$ the claim follows directly from their definitions.
  For $SU$ the claim then follows compositionally.
\end{proof*}

Now we can show:
\begin{proposition}
\label{prop:SU-properties}
Let $n \geq 1$.
\begin{enumerate}
\item \label{prop:SU-properties:(SU)*=S*U}
  $(SU)^n = S^nU$
\item \label{prop:SU-properties:absorbing}
  $SU$ is an absorbing MC with absorbing states $\set{(\emptyset, b)}{b\subseteq\Pk}$.
\end{enumerate}
\end{proposition}
\begin{proof*}~
\begin{enumerate}[leftmargin=*]
\item It suffices to show that $USU = SU$. Suppose that
\[\prob{(a,b) \reaches{USU}_1 (a',b')} = p > 0. \]
It suffices to show that this implies \[
  \prob{(a,b) \reaches{SU}_1 (a',b')} = p.
\]
If $(a,b)$ is saturated, then we must have $(a',b') = (\emptyset,b)$ and \[
  \prob{(a,b) \reaches{USU}_1 (\emptyset,b)} = 1 =
    \prob{(a,b) \reaches{SU}_1 (\emptyset,b)}
\]
If $(a,b)$ is not saturated, then $(a,b) \reaches{U}_1 (a,b)$ with
probability $1$ and therefore \[
\prob{(a,b) \reaches{USU}_1 (a',b')} = \prob{(a,b) \reaches{SU}_1 (a',b')}
\]
\item Since $S$ and $U$ are stochastic, clearly $SU$ is a MC. Since $SU$ is finite
state, any state can reach an absorbing communication class.
(To see this, note that the reachability relation $\reaches{SU}$ induces a
partial order on the communication classes of $SU$. Its maximal elements are
necessarily absorbing, and they must exist because the state space is finite.)
It thus suffices to show that a state set $C \subseteq \pset\Pk \times \pset\Pk$
in $SU$ is an absorbing communication class iff
$C = \sset{(\emptyset,b)}$ for some $b\subseteq\Pk$.
  \begin{enumerate}[leftmargin=*]
  \item[``$\Leftarrow$'':]
    Observe that $\emptyset \reaches{B}_1 a'$ iff $a'=\emptyset$. Thus
    $(\emptyset,b) \reaches{S}_1 (a',b')$ iff $a'=\emptyset$ and $b'=b$,
    and likewise $(\emptyset,b) \reaches{U}_1 (a',b')$ iff
    $a'=\emptyset$ and $b'=b$.
    Thus $(\emptyset,b)$ is an absorbing state in $SU$ as required.
  \item[``$\Rightarrow$'':]
  First observe that by monotonicity of $SU$ (\Cref{lem:monotone-chains}),
  we have $b=b'$ whenever
  $(a,b) \communicate{SU} (a',b')$; thus there exists a fixed $b_C$ such that
  $(a,b) \in C$ implies $b=b_C$.

  Now pick an arbitrary state $(a,b_C) \in C$. It suffices to show that
  $(a,b_C) \reaches{SU}(\emptyset,b_C)$, because that implies
  $(a,b_C) \communicate{SU}(\emptyset,b_C)$, which in turn implies $a=\emptyset$.
  But the choice of $(a,b_C) \in C$ was arbitrary, so that would mean
  $C = \sset{(\emptyset,b_C)}$ as claimed.

  To show that $(a,b_C) \reaches{SU} (\emptyset, b_C)$, pick arbitrary states
  such that \[
    (a,b_C) \reaches{S} (a',b') \reaches{U}_1 (a'',b'')
  \]
  and recall that this implies $(a,b_C) \reaches{SU} (a'',b'')$ by claim
  \eqref{prop:SU-properties:(SU)*=S*U}.
  Then $(a'',b'') \reaches{SU} (a,b_C)$ because $C$ is absorbing, and
  thus $b_C=b'=b''$ by monotonicity of $S$, $U$, and $SU$. But $(a',b')$ was
  chosen as an arbitrary state $S$-reachable from $(a,b_C)$, so
  $(a,b)$ and by transitivity $(a',b')$ must be saturated.
  Thus $a''=\emptyset$ by the definition of $U$. \qedhere
  \end{enumerate}
\end{enumerate}
\end{proof*}

Arranging the states $(a,b)$ in lexicographically ascending order according to
$\subseteq$ and letting $n=|\pset\Pk|$, it then follows from
\Cref{prop:SU-properties}.\ref{prop:SU-properties:absorbing} that $SU$ has the form
\begin{equation*}
  SU = \begin{bmatrix}
    I_n & 0\\
    R              & Q
  \end{bmatrix}
\end{equation*}
where for $a \neq \emptyset$
\begin{equation*}
  (SU)_{(a,b),(a',b')} = \begin{bmatrix}R&Q\end{bmatrix}_{(a,b),(a',b')}
\end{equation*}
Moreover, $SU$ converges and its limit is given by
\begin{equation}\label{eq:SU-limit-closed-form}
  (SU)^\infty \defeq \begin{bmatrix}
    I_n           & 0\\
    (I-Q)^{-1}R   & 0
  \end{bmatrix}
  = \lim_{n\to\infty} (SU)^n
\end{equation}
We can use the modified Markov chain $SU$ to compute the limit of $S$:
\begin{theorem}[Closed Form]
\label{thm:closed-form}
Let $a,b,b' \subseteq \Pk$. Then
\begin{equation}\label{eq:limit-of-S}
  \lim_{n\to\infty} \sum_{a'} S^n_{(a,b),(a',b')} = (SU)^\infty_{(a,b),(\emptyset,b')}
\end{equation}
or, using matrix notation,
\begin{equation}\label{eq:limit-of-S-matrix-form}
  \lim_{n\to\infty} \sum_{a'} S^n_{(-,-),(a',-)} = \begin{bmatrix}
    I_n\\
    (I-Q)^{-1}R
  \end{bmatrix} \in [0,1]^{(\pset\Pk \times \pset\Pk) \times \pset\Pk}
\end{equation}
In particular, the limit in \cref{eq:limit-of-S} exists and it can be
effectively computed in closed-form.
\end{theorem}
\begin{proof*}
Using \Cref{prop:SU-properties}.\ref{prop:SU-properties:(SU)*=S*U} in the second
step and equation \cref{eq:SU-limit-closed-form} in the last step,
\begin{align*}
\lim_{n\to\infty} \sum_{a'} S^n_{(a,b),(a',b')}
&= \lim_{n\to\infty} \sum_{a'} (S^nU)_{(a,b),(a',b')}\\
&= \lim_{n\to\infty} \sum_{a'} (SU)^n_{(a,b),(a',b')}\\
&= \sum_{a'} (SU)^\infty_{(a,b),(a',b')} = (SU)^\infty_{(a,b),(\emptyset,b')}
\end{align*}
$(SU)^\infty$ is computable because $S$ and $U$ are matrices over $\Q$ and hence
so is $(I-Q)^{-1}R$.
\end{proof*}

\begin{corollary}
\label{cor:equiv-decidable}
For programs $p$ and $q$, it is decidable whether $\polp \equiv \polq$.
\end{corollary}
\begin{proof*}
Recall from \Cref{cor:den-equal-iff-bden-equal} that it suffices to compute the
finite rational matrices $\bden{\polp}$ and $\bden{\polq}$ and check them for equality.
But \Cref{thm:closed-form} together with \Cref{prop:small-step-characterization}
gives us an effective mechanism to compute $\bden{-}$ in the case of Kleene star,
and $\bden{-}$ is straightforward to compute in all other cases.

To summarize, we repeat the full chain of equalities we have deduced:
\begin{equation*}
  \den{\polp\star}(a)(\sset{b}) = \bden{\polp\star}_{a,b}
  = \lim_{n\to\infty} \bden{\polp^{(n)}}_{a,b}
  = \lim_{n\to\infty} \sum_{a'} \sden{\polp}^n_{(a,\emptyset),(a',b)}
  = (SU)^\infty_{(a,\emptyset),(\emptyset,b)}
\end{equation*}
(From left to right: \Cref{thm:big-step-sound}, Definition of $\bden{-}$, \Cref{prop:small-step-characterization}, and \Cref{thm:closed-form}.)
\end{proof*}

\section{Case Study: Resilient Routing}
\label{sec:case}


We have build a prototype based on \Cref{thm:closed-form} and
\Cref{cor:equiv-decidable} in OCaml. It implements \probnetkat as an embedded
DSL and compiles \probnetkat programs to transition matrices using
symbolic techniques and a sparse linear algebra solver. A detailed
description and performance evaluation of the implementation is beyond
the scope of this paper. Here we focus on demonstrating the utility of
such a tool by performing a case study with real-world datacenter
topologies and resilient routing schemes.

Recently proposed datacenter
designs~\cite{liu2013f10,niranjan2009portland,al2008scalable,singla2012jellyfish,guo2009bcube,guo2008dcell}
utilize a large number of inexpensive commodity switches, which
improves scalability and reduces cost compared to other approaches.
However, relying on many commodity devices also increases the
probability of failures. A recent measurement study showed that
network failures in datacenters~\cite{gill11} can have a major impact
on application-level performance, leading to a new line of work
exploring the design of fault-tolerant datacenter fabrics. Typically
the topology and routing scheme are co-designed, to achieve good
resilience while still providing good performance in terms of
throughput and latency.


\begin{figure}
\centering
  \begin{minipage}[b]{0.44\textwidth}
    \input{img/fattree}
    \caption{A FatTree topology with 20 switches.}
    \label{fig:fattree}
  \end{minipage}\hspace{0pt}%
  \begin{minipage}[b]{0.54\textwidth}
    \input{img/abfattree}
    \caption{An AB FatTree topology with 20 switches.}
    \label{fig:abfattree}
  \end{minipage}%
\end{figure}%

\subsection{Topology and routing}

Datacenter topologies typically organize the fabric into multiple
levels of switches.

\paragraph{FatTree.}%
A FatTree~\cite{al2008scalable}, which is a multi-level, multi-rooted
tree, is perhaps the most common example of such a
topology. ~\Cref{fig:fattree} shows a 3-level FatTree topology with 20
switches. The bottom level,
\emph{edge}, consists of top-of-rack (ToR) switches; each ToR
switch connects all the hosts within a rack (not shown in the figure).
These switches act as
ingress and egress for intra-datacenter traffic. The other two
levels, \emph{aggregation} and \emph{core}, redundantly interconnect
the switches from the edge layer.

The redundant structure of a FatTree naturally lends itself to
forwarding schemes that locally route around failures.  To illustrate,
consider routing from a source ($s7$) to a destination $(s1)$ along
shortest paths in the example topology.  Packets are first forwarded
upwards, until eventually there exists a downward path to $s1$. The
green links in the figure depict one such path. On the way up, there
are multiple paths at each switch that can be used to forward
traffic. Thus, we can route around failures by simply choosing an
alternate upward link. A common routing scheme is called equal-cost
multi-path routing (ECMP) in the literature, because it chooses
between several paths all having the same \emph{cost}---\eg, path
length. ECMP is especially attractive as is it can provide better
resilience without increasing the lengths of forwarding paths.

However, after reaching a core switch, there is a \emph{unique}
shortest path to the destination, so ECMP no longer provides any
resilience if a switch fails in the aggregation layer (\cf the red
cross in \Cref{fig:fattree}). A more sophisticated scheme could take a
longer (5-hop) detour going all the way to another edge switch, as
shown by the red lines in the figure. Unfortunately, such detours
inflate the path length and lead to increased latency and congestion.

\paragraph{AB FatTree}%
FatTree's unpleasantly long backup routes on the downward paths are
caused by the symmetric wiring of aggregation and core switches. AB
FatTrees~\cite{liu2013f10} alleviate this flaw by skewing the symmetry
of wiring. It defines two types of subtrees, differing in their wiring
to higher levels. To illustrate, ~\cref{fig:abfattree} shows an
example which rewires the FatTree from ~\cref{fig:fattree} to make it
an AB FatTree. It contains two types of subtrees:
\begin{enumerate}[label={\textit{\roman*})}]
  \item{Type A:} switches depicted in blue and wired to core using dashed lines, and
  \item{Type B:} switches depicted in red and wired to core using solid lines.
\end{enumerate}
Type A subtrees are wired in a way similar to FatTree, but type B
subtrees differ in their connections to core switches (see the
original paper for full details~\cite{liu2013f10}).

This slight change in wiring enables shorter detours to route around
failures in the downward direction. Consider again a flow involving
the same source ($s7$) and destination ($s1$). As before, we have
multiple options going upwards when following shortest paths (\eg, the
one depicted in green), but we have a unique downward path once we
reach the top. But unlike FatTree, if the aggregation switch on the
downward path fails, we find that there is a short (3-hop) detour, as
shown in blue. This backup path exists because the core switch, which
needs to reroute traffic, is connected to aggregation switches of both
types of subtrees. More generally, aggregation switches of the same
type as the failed switch provide a 5-hop detour (as in a standard
FatTrees); but aggregation switches of the opposite type can provide a
more efficient 3-hop detour.

\subsection{\probnetkat implementation.}

\begin{figure}
\centerline{\includegraphics[width=\textwidth,draft=false]{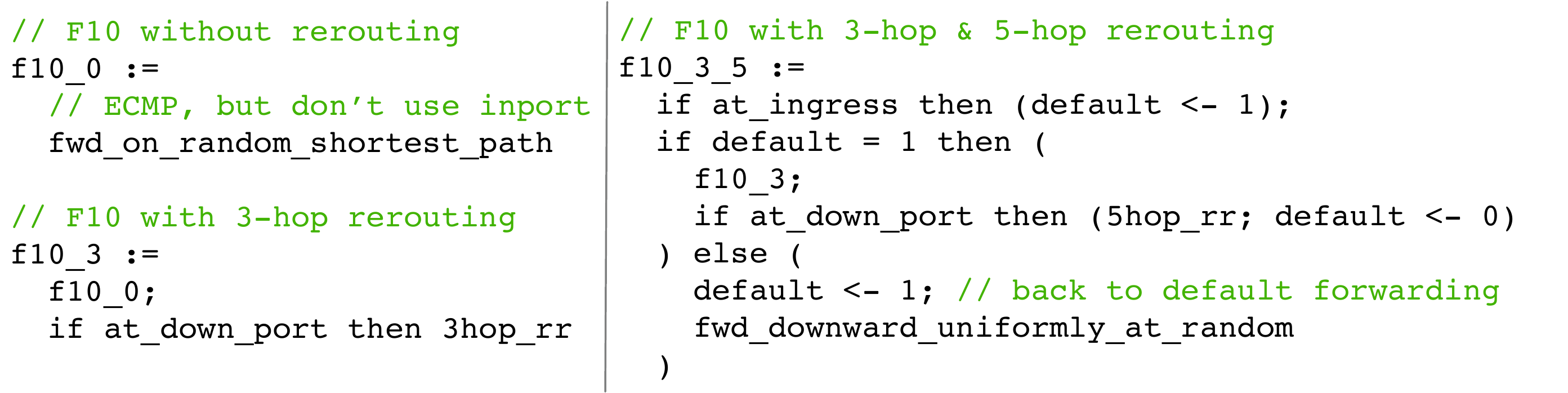}}
\vspace{-1ex}
\caption{\probnetkat implementation of F10 in three refinement steps.}
\label{fig:f10-code}
\end{figure}

Now we will see how to encode several routing schemes
using \probnetkat and analyze their behavior in each topology under
various failure models.

\paragraph{Routing}%
\ften{} \cite{liu2013f10} provides a routing algorithm that combines
the three routing and rerouting strategies we just discussed (ECMP,
3-hop rerouting, 5-hop rerouting) into a single scheme.  We
implemented it in three steps (see \Cref{fig:f10-code}).  The first
scheme, \ften{0}, implements an approach similar to
ECMP:\footnote{ECMP implementations are usually based on hashing,
which approximates random forwarding provided there is sufficient
entropy in the header fields used to select an outgoing port.} it
chooses a port uniformly at random from the set of ports connected to
minimum-length paths to the destination. We exclude the port at which
the packet arrived from this set; this eliminates the possibility of
forwarding loops when routing around failures.

Next, we improve the resilience of \ften{0} by augmenting it with
3-hop rerouting if the next hop aggregation switch $A$ along the
downward shortest path from a core switch $C$ fails. To illustrate,
consider the blue path in \Cref{fig:abfattree}. We find a port on $C$
that connects to an aggregation switch $A'$ of the opposite type than
the failed aggregation switch, $A$, and forward the packet to $A'$. If
there are multiple such ports that have not failed, we choose one
uniformly at random. Normal routing continues at $A'$, and ECMP will
know not to send the packet back to $C$. \ften{3} implements this
refinement.

Note that if the packet is still parked at port whose adjacent link is
down after executing $\ften{3}$, it must be that all ports connecting
to aggregation switches of the opposite type are down.  In this case,
we attempt 5-hop rerouting via an aggregation switch $A''$ of the same
type as $A$. To illustrate, consider the red path
in \Cref{fig:abfattree}. We begin by sending the packet to $A''$. To
let $A''$ know that it should not send the packet back up as normally,
we set a flag $\mathtt{default}$ to false in the packet, telling $A''$
to send the packet further down instead.  From there, default routing
continues. \ften{3,5} implements this refinement. 

\begin{table}
\centering
  \begin{minipage}[b]{0.46\textwidth}
\centering
\tabcolsep=0.11cm
\begin{tabular}{lccc}
\toprule
$k$ & \thead{$\hat\model(\ften{0}, t, f_k)$ \\ $\equiv {\mathit{teleport}}$}
  & \thead{$\hat\model(\ften{3}, t, f_k)$ \\ $ \equiv {\mathit{teleport}}$}
  & \thead{$\hat\model(\ften{3,5}, t, f_k)$ \\ $ \equiv {\mathit{teleport}}$} \\
\midrule
0 & \cmark & \cmark & \cmark \\
1 & \xmark & \cmark & \cmark \\
2 & \xmark & \cmark & \cmark \\
3 & \xmark & \xmark & \cmark \\
4 & \xmark & \xmark & \xmark \\
$\infty$ & \xmark & \xmark & \xmark \\
\bottomrule
\end{tabular}
\caption{Evaluating $k$-resilience of F10.}
\label{tab:equiv-teleport}

  \end{minipage}\hspace{0pt}%
  \begin{minipage}[b]{0.54\textwidth}

\centering
\tabcolsep=0.11cm
\begin{tabular}{lccc}
\toprule
$k$ & \thead{$\mathtt{compare}$ \\ $(\ften{0}, \ften{3}\mathtt{)}$}
  & \thead{$\mathtt{compare}$ \\ $(\ften{3}, \ften{3,5}\mathtt{)}$}
  & \thead{$\mathtt{compare}$ \\ $(\ften{3,5}, \mathit{teleport}\mathtt{)}$}\\
\midrule
0 & $\equiv$ & $\equiv$ & $\equiv$ \\
1 & $<$ & $\equiv$ & $\equiv$ \\
2 & $<$ & $\equiv$ & $\equiv$ \\
3 & $<$ & $<$ & $\equiv$ \\
4 & $<$ & $<$ & $<$ \\
$\infty$ & $<$ & $<$ & $<$ \\
\bottomrule
\end{tabular}
\caption{Comparing schemes under $k$ failures.}
\label{tab:comp-scheme}
\end{minipage}%
\end{table}%

\paragraph{Failure and Network model}%
We define a family of failure models $f^p_k$ in the style
of \Cref{sec:overview}. Let $k \in \N \cup \sset{\infty}$ denote a
bound on the maximum number of link failures that may occur
simultaneously, and assume that links otherwise fail independently
with probability $0 \leq p < 1$ each. We omit $p$ when it is clear
from context. For simplicity, to focus on the more complicated
scenarios occurring on downward paths, we will model failures only for
links connecting the aggregation and core layer.

Our network model works much like the one from \Cref{sec:overview}.
However, we model a single destination, switch 1, and we elide the
final hop to the appropriate host connected to this switch.
\[
  \model(p,t) \defeqs \pseq{in}{\dowhile{(\pseq{p}{t})}{(\pnot{\match{\sw}{1}})}}
\]
The ingress predicate $in$ is a disjunction of switch-and-port tests
over all ingress locations. This first model is embedded into a
refined model $\hat\model(p,t,f)$ that integrates the failure model
and declares all necessary local variables that track the healthiness
of individual ports:
\begin{align*}
\hat\model(p,t,f) \defeqs~
  &\kw{var}~\modify{\up{1}}{1}~\kw{in}\\
  &\quad\dots\\
  &\kw{var}~\modify{\up{d}}{1}~\kw{in}\\
  &\model((\pseq{f}{p}), t)
\end{align*}
Here $d$ denotes the maximum degree of all nodes in the FatTree and AB
FatTree topologies from \Cref{fig:fattree,fig:abfattree}, which we
encode as programs $\mathit{fattree}$ and $\mathit{abfattree}$.
much like in \Cref{subsec:overview-prob}.


\subsection{Checking invariants}
We can gain confidence in the correctness of our implementation
of \ften{} by verifying that it maintains certain key invariants. As
an example, recall our implementation of \ften{3,5}: when we perform
5-hop rerouting, we use an extra bit ($\mathit{default}$) to notify
the next hop aggregation switch to forward the packet downwards
instead of performing default forwarding. The next hop follows this
instruction and also sets $\mathit{default}$ back to $1$. By design,
the packet can not be delivered to the destination with
$\mathit{default}$ set to $0$. 

To verify this property, we check the following equivalence:
\[
\forall t,k: \ \ \hat\model(\ften{3,5}, \mathit{t}, f_k) \ \equiv \
\pseq{\hat\model(\ften{3,5}, \mathit{t}, f_k)}
\match{\mathit{default}}{1}
\]
We executed the check using our implementation for $k \in \{0, 1, 2,
3, 4, \infty \}$ and
$t \in \sset{\mathit{fattree}, \mathit{abfattree}}$. As discussed
below, we actually failed to implement this feature correctly on our
first attempt due to a subtle bug---we neglected to initialize the
$\mathtt{default}$ flag to $1$ at the ingress.

\subsection{F10 routing with FatTree}
We previously saw that the structure of FatTree doesn't allow 3-hop
rerouting on failures because all subtrees are of the same type. This
would mean that augmenting ECMP with 3-hop rerouting should have no
effect, \ie 3-hop rerouting should never kick in and act as a no-op.
To verify this, we can check the following equivalence:
\[
\forall k: \ \ \hat\model(\ften{0}, \mathit{fattree}, f_k) \ \equiv \
\hat\model(\ften{3}, \mathit{fattree}, f_k)
\]
We have used our implementation to check that this equivalence indeed
holds for $k \in \{0, 1, 2, 3, 4, \infty \}$.

\subsection{Refinement}
Recall that we implemented \ften{} in three stages. We started with a
basic routing scheme (\ften{0}) based on ECMP that provides resilience
on the upward path, but no rerouting capabilities on the downward
paths. We then augmented this scheme by adding 3-hop rerouting to
obtain \ften{3}, which can route around certain failures in the
aggregation layer. Finally, we added 5-hop rerouting to address
failure cases that 3-hop rerouting cannot handle,
obtaining \ften{3,5}. Hence, we would expect the probability of packet
delivery to increase with each refinement of our routing
scheme. Additionally, we expect all schemes to deliver packets and
drop packets with some probability under the unbounded failure model.
These observations are summarized by the following ordering:
\[
  \pfalse \ <\ \hat\model(\ften{0}, t, f_{\infty}) \ < \ \hat\model(\ften{3}, t, f_{\infty})
  \ < \ \hat\model(\ften{3,5}, t, f_{\infty}) \ < \ {\mathit{teleport}}
\]
where $t=\mathit{abfattree}$ and $\mathit{teleport} \defeq \modify{\sw}{1}$.
To our surprise, we were not able to verify this property initially,
as our implementation indicated that the ordering
\[
\hat\model(\ften{3}, t, f_{\infty})
  \ < \ \hat\model(\ften{3,5}, t, f_{\infty})
\]
was violated. We then added a capability to our implementation to
obtain counterexamples, and found that $\ften{3}$ performed better
than $\ften{3,5}$ for packets $\pk$ with $\pk.\mathit{default} =
0$. We were missing the first line in our implementation of
$\ften{3,5}$ (\cf, \Cref{fig:f10-code}) that initializes the
$\mathit{default}$ bit to 1 at the ingress, causing packets to be
dropped! After fixing the bug, we were able to confirm the expected
ordering.

\begin{figure}[t!]
  \includegraphics[width=.65\textwidth,draft=false]{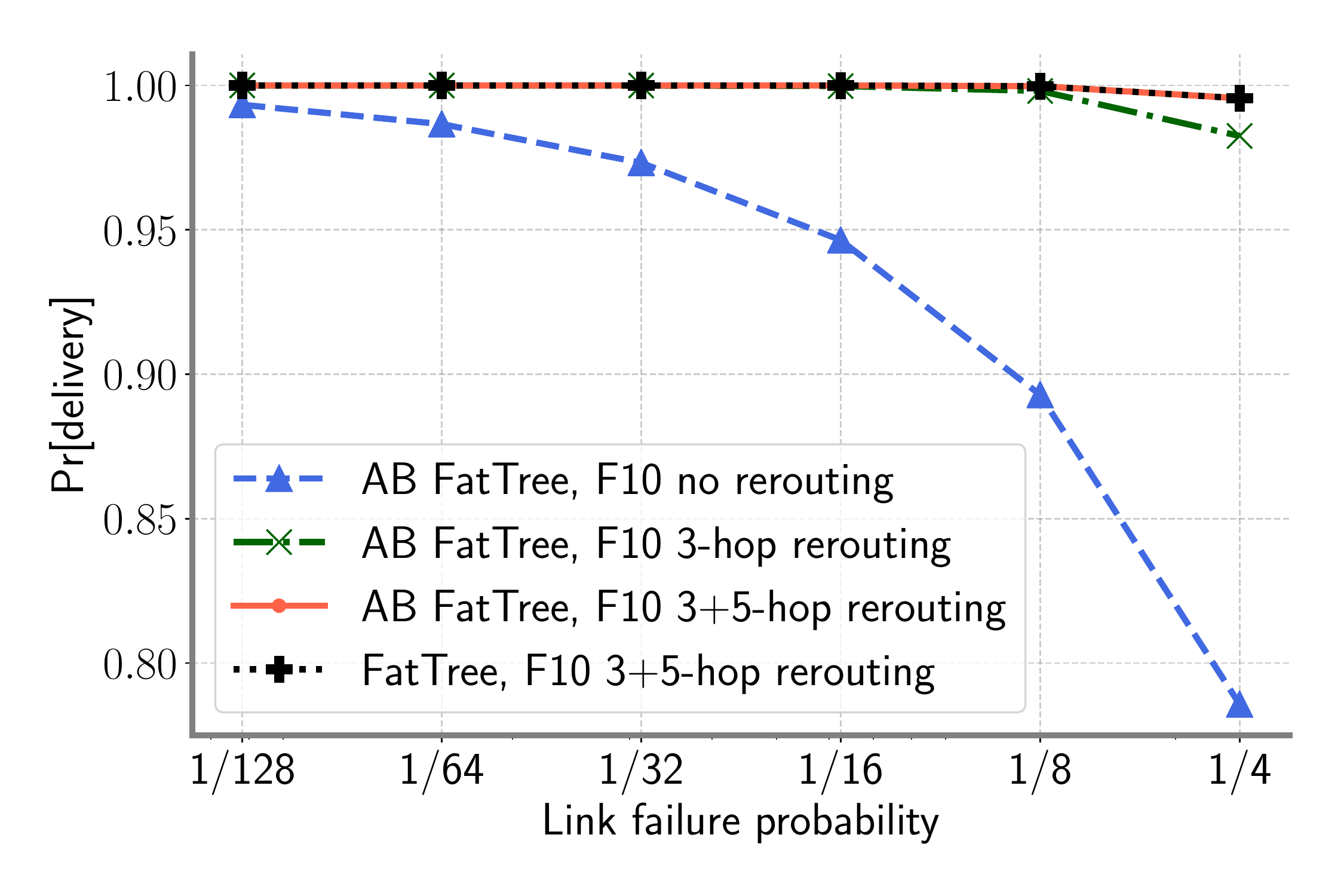}
  \caption{Probability of delivery vs. link-failure probability.  ($k=\infty$).}
  \label{fig:tput-vs-fp}
\end{figure}

\subsection{k-resilience}
We saw that there exists a strict ordering in terms of resilience for
\ften{0}, \ften{3} and \ften{3,5} when an unbounded number of failures can
happen. Another interesting way of measuring resilience is to count
the minimum number of failures at which a scheme fails to guarantee
100\% delivery. Using ProbNetKAT, we can measure this resilience by
setting $k$ in $f_k$ to increasing values and checking equivalence
with teleportation. \Cref{tab:equiv-teleport} shows the results based
on our decision procedure for the AB FatTree topology from
~\Cref{fig:abfattree}.

The naive scheme, \ften{0}, which does not perform any rerouting,
drops packets when a failure occurs on the downward path. Thus, it is
0-resilient. In the example topology, 3-hop rerouting has two possible
ways to reroute for the given failure. Even if only one of the type B
subtrees is reachable, \ften{3} can still forward traffic. However, if
both the type B subtrees are unreachable, then \ften{3} will not be
able to reroute traffic. Thus, \ften{3} is 2-resilient. Similarly,
\ften{3,5} can route as long as any aggregation switch is
reachable from the core switch. For \ften{3,5} to fail the core switch
would need to be disconnected from all four aggregation
switches. Hence it is 3-resilient. In cases where schemes are not
equivalent to $\mathit{teleport}$, we can characterize the relative
robustness by computing the ordering, as shown in
~\Cref{tab:comp-scheme}.

\subsection{Resilience under increasing failure rate}

We can also do more quantitative analyses such as evaluating the
effect of increase failure probability of links on the probability of
packet delivery. \Cref{fig:tput-vs-fp} shows this analysis in a
failure model in which an unbounded number of failures can occur
simultaneously. We find that \ften{0}'s delivery probability dips
significantly as the failure probability increases because \ften{0} is
not resilient to failures. In contrast, both \ften{3} and \ften{3,5}
continue to ensure high probability of delivery by rerouting around
failures.

\begin{figure}[t!]
  \centering
  \includegraphics[width=.65\textwidth,draft=false]{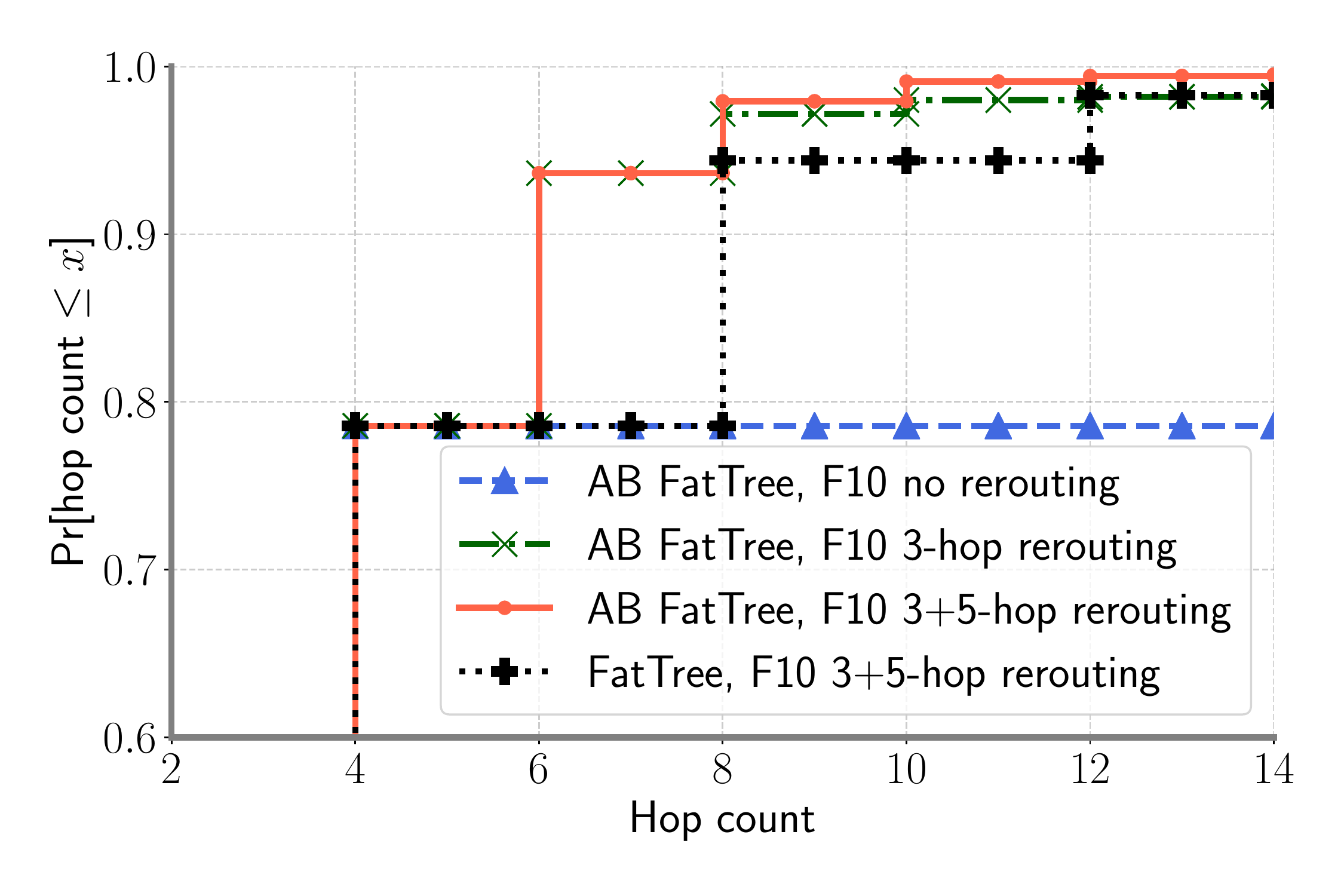}
  \caption{Increased latency due to resilience. ($k=\infty$, $p=\frac{1}{4}$)}
  \label{fig:hop-count-cdf}
\end{figure}

\begin{figure}[t!]
  \includegraphics[width=.65\textwidth,draft=false]{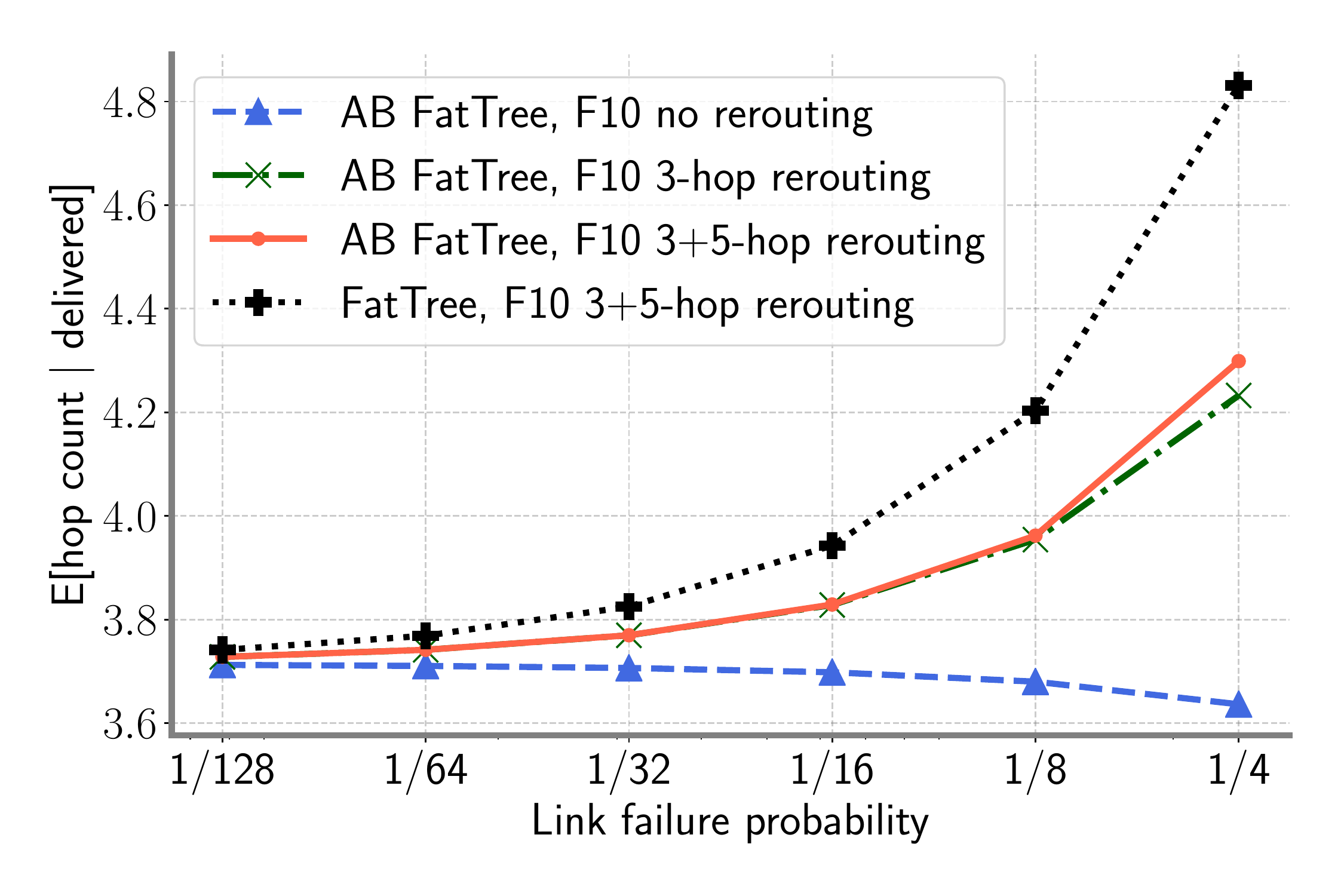}
  \caption{Expected hop-count conditioned on delivery.  ($k=\infty$).}
  \label{fig:expected-hop-count}
\end{figure}

\subsection{Cost of resilience}
By augmenting naive routing schemes with rerouting mechanisms, we are
able to achieve a higher degree of resilience. But this benefit comes
at a cost. The detours taken to reroute traffic increase the latency
(hop count) for packets. \probnetkat enables quantifying this increase
in latency by augmenting our model with a counter that gets increased
at every hop. \Cref{fig:hop-count-cdf} shows the CDF of latency as the
fraction of traffic delivered within a given hop count. On AB FatTree,
we find that \ften{0} delivers as much traffic as it can
($\approx$80\%) within a hop count $\leq$ 4 because the maximum length
of a shortest path from any edge switch to $s1$ is 4 and \ften{0} does
not use any longer paths. \ften{3} and \ften{3,5} deliver the same
amount of traffic with hop count $\leq 4$. But, with 2 additional
hops, they are able to deliver significantly more traffic because they
perform 3-hop rerouting to handle certain failures. With 4 additional
hops, \ften{3,5}'s throughput increases as 5-hop rerouting helps. We
find that \ften{3} also delivers more traffic with 8 hops---these are
the cases when \ften{3} performs 3-hop rerouting twice for a single
packet as it encountered failure twice. Similarly, we see small
increases in throughput for higher hop counts. We find that \ften{3,5}
improves resilience for FatTree too, but the impact on latency is
significantly higher as FatTree does not support 3-hop rerouting.

\subsection{Expected latency}

\cref{fig:expected-hop-count} shows the expected hop-count of paths
taken by packets conditioned on their delivery. Both \ften{3} and
\ften{3,5} deliver packets with high probability even at high failure
probabilities, as we saw in \Cref{fig:tput-vs-fp}. However, a higher
probability of link-failure implies that it becomes more likely for
these schemes to invoke rerouting, which increases hop count.  Hence,
we see the increase in expected hop-count as failure probability
increases. \ften{3,5} uses 5-hop rerouting to achieve more resilience
compared to \ften{3}, which performs only 3-hop rerouting, and this
leads to slightly higher expected hop-count for \ften{3,5}. We see
that the increase is more significant for FatTree in contrast to AB
FatTree because FatTree only supports 5-hop rerouting.

As the failure probability increases, the probability of delivery for
packets that are routed via the core layer decreases significantly
for \ften{0} (recall \Cref{fig:tput-vs-fp}).  Thus, the distribution
of delivered packets shifts towards those with direct 2-hop path via
an aggregation switch (such as packets from $s2$ to $s1$), and hence
the expected hop-count decreases slightly.

\subsection{Discussion}

As this case study of resilient routing in datacenters shows, the
stochastic matrix representation of \probnetkat programs and
accompanying decision procedure enable us to answer a wide variety of
questions about probabilistic networks completely automatically. These
new capabilities represent a signficant advance over current network
verification tools, which are based on deterministic packet-forwarding
models~\cite{hsa,anteater,veriflow,FKMST15a}.

\section{Deciding Full \probnetkat: Obstacles and Challenges}
\label{sec:full}

As we have just seen, history-free \probnetkat can describe
sophisticated network routing schemes under various failure models,
and program equivalence for the language is decidable. However, it is
less expressive than the original \probnetkat language, which includes
an additional primitive $\pdup$. Intuitively, this command duplicates
a packet $\pk \in \Pk$ and outputs the word $\pk\pk \in \Hist$, where
$\Hist = \Pk^*$ is the set of non-empty, finite sequences of
packets. An element of $\Hist$ is called a \emph{packet history},
representing a log of previous packet states. \probnetkat policies may
only modify the first (\emph{head}) packet of each history; $\pdup$
fixes the current head packet into the log by copying it. In this
way, \probnetkat policies can compute distributions over the paths
used to forward packets, instead of just over the final output
packets.

However, with $\pdup$, the semantics of \probnetkat becomes
significantly more complex.  Policies $\polp$ now transform sets of
packet histories $a \in \pH$ to distributions
$\den{\polp}(a) \in \Dist(\pH)$. Since $\pH$ is uncountable, these
distributions are no longer guaranteed to be discrete, and formalizing
the semantics requires full-blown measure theory (see prior work for
details~\cite{probnetkat-scott}).

Deciding program equivalence also becomes more challenging. Without
$\pdup$, policies operate on sets of packets $\pPk$; crucially, this
is a \emph{finite} set and we can represent each set with a single
state in a finite Markov chain. With $\pdup$, policies operate on sets
of packet histories $\pH$.  Since this set is not finite---in fact, it
is not even countable---encoding each packet history as a state would
give a Markov chain with infinitely many states. Procedures for
deciding equivalence are not known for such systems.

While in principle there could be a more compact representation of
general \probnetkat policies as finite Markov chains or other models
where equivalence is decidable, (\eg, weighted or probabilistic
automata~\cite{droste2009handbook} or quantitative variants of regular
expressions~\cite{alur2016regular}), we suspect that deciding
equivalence in the presence of $\dup$ is intractable. As evidence in
support of this conjecture, we show that \probnetkat policies can
simulate the following kind of probabilistic automata. This model
appears to be new, and may be of independent interest.

\begin{definition}
\label{def:bi-auto}
Let $A$ be a finite alphabet. A \emph{2-generative probabilistic
automata} is defined by a tuple $(S, s_0, \rho, \tau)$ where $S$ is a
finite set of states; $s_0 \in S$ is the initial state; $\rho : S \to
(A \cup \{ \blank \})^2$ maps each state to a pair of letters $(u,
v)$, where either $u$ or $v$ may be a special blank character
$\blank$; and the transition function $\tau : S \to
\Dist(S)$ gives the probability of transitioning from one state to another.
\end{definition}

The semantics of an automaton can be defined as a probability measure
on the space $A^\infty \times A^\infty$, where $A^\infty$ is the set
of finite and (countably) infinite words over the alphabet
$A$. Roughly, these measures are fully determined by the probabilities
of producing any two finite prefixes of words $(w, w') \in A^* \times
A^*$.

Presenting the formal semantics would require more concepts from
measure theory and take us far afield, but the basic idea is simple to
describe. An infinite trace of a 2-generative automata over states
$s_0, s_1, s_2, \dots$ gives a sequence of pairs of (possibly blank)
letters:
\[
\rho(s_0), \rho(s_1), \rho(s_2) \dots
\]
By concatenating these pairs together and dropping all blank
characters, a trace induces two (finite or infinite) words over the
alphabet $A$. For example, the sequence,
\[
(a_0,\blank), (a_1, \blank), (\blank, a_2), \dots
\]
gives the words $a_0 a_1 \dots$ and $a_2 \dots$. Since the traces are
generated by the probabilistic transition function $\tau$, each
automaton gives rise to a probability measure over pairs of words.

While we have no formal proof of hardness, deciding equivalence
between these automata appears highly challenging. In the special case
where only one word is generated (say, when the second component
produced is always blank), these automata are equivalent to standard
automata with $\epsilon$-transitions (\eg,
see \cite{mohri2000generic}).  In the standard setting, non-productive
steps can be eliminated and the automata can be modeled as a finite
state Markov chain, where equivalence is decidable. In our setting,
however, steps producing blank letters in one component may produce
non-blank letters in the other. As a result, it is not entirely clear
how to eliminate these steps and encode our automata as a Markov
chain.

Returning to \probnetkat, 2-generative automata can be encoded as
policies with $\pdup$. We sketch the idea here, deferring further
details to Appendix~\ref{app:bi-auto}. Suppose we are given an
automaton $(S, s_0, \rho, \tau)$. We build a \probnetkat policy over
packets with two fields, $\kw{st}$ and $\kw{id}$. The first field
$\kw{st}$ ranges over the states $S$ and the alphabet $A$, while the
second field $\kw{id}$ is either $1$ or $2$; we suppose the input set
has exactly two packets labeled with $\kw{id} = 1$ and $\kw{id} = 2$.
In a set of packet history, the two active packets have the same value
for $\kw{st} \in S$---this represents the current state in the
automata. Past packets in the history have $\kw{st} \in A$,
representing the words produced so far; the first and second
components of the output are tracked by the histories with $\kw{id} =
1$ and $\kw{id} = 2$. We can encode the transition function $\tau$ as
a probabilistic choice in \probnetkat, updating the current state
$\kw{st}$ of all packets, and recording non-blank letters produced by
$\rho$ in the two components by applying $\pdup$ on packets with the
corresponding value of $\kw{id}$.

Intuitively, a set of packet histories generated by the
resulting \probnetkat term describes a pair of words generated by the
original automaton.  With a bit more bookkeeping (see
Appendix~\ref{app:bi-auto}), we can show that two 2-generative
automata are equivalent if and only if their encoded \probnetkat
policies are equivalent. Thus, deciding equivalence for \probnetkat
with $\pdup$ is harder than deciding equivalence for 2-generative
automata. Showing hardness for the full framework is a fascinating
open question. At the same time, deciding equivalence between
$2$-generative automata appears to require substantially new ideas;
these insights could shed light on how to decide equivalence for the
full \probnetkat language.

\section{Related Work}
\label{sec:rw}

A key ingredient that underpins the results in this paper is the idea
of representing the semantics of iteration using absorbing Markov
chains, and exploiting their properties to directly compute limiting
distributions on them. 

Markov chains have been used by several authors to represent and to
analyze probabilistic programs. 
%
An early example of using Markov chains for modeling probabilistic
programs is the seminal paper by Sharir, Pnueli, and
Hart \cite{sharir1984verification}. They present a general method for
proving properties of probabilistic programs. In their work, a
probabilistic program is modeled by a Markov chain and an assertion on
the output distribution is extended to an invariant assertion on all
intermediate distributions (providing a probabilistic generalization
of Floyd's inductive assertion method).
Their approach can assign semantics to infinite Markov chains for
infinite processes, using stationary distributions of absorbing Markov
chains in a similar way to the one used in this paper. Note however
that the state space used in this and other work is not
like \probnetkat's current and accumulator sets ($2^{Pk} \times
2^{Pk}$), but is instead is the Cartesian product of variable
assignments and program location. In this sense, the absorbing states
occur for program termination, rather than for accumulation as
in \probnetkat. Although packet modification is clearly related to
variable assignment, accumulation does not clearly relate to program
location.

Readers familiar with prior work on probabilistic automata might
wonder if we could directly apply known results on (un)decidability of
probabilistic rational languages. This is not the case---probabilistic
automata accept distributions over words, while \probnetkat programs
encode distributions over languages. Similarly, probabilistic
programming languages, which have gained popularity in the last decade
motivated by applications in machine learning, focus largely on
Bayesian inference. They typically come equipped with a primitive for
probabilistic conditioning and often have a semantics based on
sampling. Working with
\probnetkat has a substantially different style, in that the focus is on 
on specification and verification rather than inference.

Di Pierro, Hankin, and Wiklicky have used probabilistic abstract
interpretation (PAI) to statically analyze probabilistic
$\lambda$-calculus \cite{di2005probabilistic}. They introduce a linear
operator semantics (LOS) and demonstrate a strictness analysis, which
can be used in deterministic settings to replace lazy with eager
evaluation without loss.
%
Their work was later extended to a language called $pWhile$, using a
store plus program location state-space similar
to \cite{sharir1984verification}.
The language $pWhile$ is a basic imperative language comprising
while-do and if-then-else constructs, but augmented with random choice
between program blocks with a rational probability, and limited to a
finite of number of finitely-ranged variables (in our case, packet
fields). The authors explicitly limit integers to finite sets for
analysis purposes to maintain finiteness, arguing that real programs
will have fixed memory limitations.
%
In contrast to our work, they do not deal with infinite limiting
behavior beyond stepwise iteration, and do not guarantee convergence.
Probabilistic abstract interpretation is a new but growing field of
research~\cite{prob-AI}.
 
Olejnik, Wicklicky, and Cheraghchi provided a probabilistic compiler
$pwc$ for a variation of $pWhile$ \cite{olejnik2016probabilistic},
implemented in OCaml, together with a testing framework.  The $pwc$
compiler has optimizations involving, for instance, the Kronecker
product to help control matrix size, and a Julia backend. Their
optimizations based on the Kronecker product might also be applied in,
for instance, the generation of $\sden{p}$ from $\bden{p}$, but we
have not pursued this direction as of yet.

There is a plenty of prior work on finding explicit distributions of
probabilistic programs. Gordon, Henzinger, Nori, and Rajamani surveyed
the state of the art with regard to probabilistic inference
\cite{gordon2014probabilistic}. They show how stationary distributions on Markov
chains can be used for the semantics of infinite probabilistic processes, and
how they converge under certain conditions. Similar to our approach, they use
absorbing strongly-connected-components to represent termination.

Markov chains are used in many probabilistic model checkers, of which
PRISM \cite{kwiatkowska2011prism} is a prime example.  PRISM supports
analysis of discrete-time Markov chains, continuous-time Markov
chains, and Markov decision processes. The models are checked against
specifications written in temporal logics like PCTL and CSL. PRISM is
written in Java and C++ and provides three model checking engines: a
symbolic one with (multi-terminal) binary decision diagrams
((MT)BDDs), a sparse matrix one, and a hybrid.
%
The use of PRISM to analyse ProbNetKAT programs is an interesting
research avenue and we intend to explore it in the future.

\section{Conclusion}
\label{sec:conc}

This paper settles the decidability of program equivalence for
history-free \probnetkat. The key technical challenge is overcome by
modeling the iteration operator as an absorbing Markov chain, which
makes it possible to compute a closed-form solution for its
semantics. The resulting tool is useful for reasoning about a host of
other program properties unrelated to equivalence. Natural directions
for future work include investigating equivalence for full
\probnetkat, developing an optimized implementation, and exploring new
applications to networks and beyond.


\bibliography{icfp}


\begin{thebibliography}{34}


\ifx \showCODEN    \undefined \def \showCODEN     #1{\unskip}     \fi
\ifx \showDOI      \undefined \def \showDOI       #1{#1}\fi
\ifx \showISBNx    \undefined \def \showISBNx     #1{\unskip}     \fi
\ifx \showISBNxiii \undefined \def \showISBNxiii  #1{\unskip}     \fi
\ifx \showISSN     \undefined \def \showISSN      #1{\unskip}     \fi
\ifx \showLCCN     \undefined \def \showLCCN      #1{\unskip}     \fi
\ifx \shownote     \undefined \def \shownote      #1{#1}          \fi
\ifx \showarticletitle \undefined \def \showarticletitle #1{#1}   \fi
\ifx \showURL      \undefined \def \showURL       {\relax}        \fi
\providecommand\bibfield[2]{#2}
\providecommand\bibinfo[2]{#2}
\providecommand\natexlab[1]{#1}
\providecommand\showeprint[2][]{arXiv:#2}

\bibitem[\protect\citeauthoryear{Al-Fares, Loukissas, and Vahdat}{Al-Fares
  et~al\mbox{.}}{2008}]%
        {al2008scalable}
\bibfield{author}{\bibinfo{person}{Mohammad Al-Fares},
  \bibinfo{person}{Alexander Loukissas}, {and} \bibinfo{person}{Amin Vahdat}.}
  \bibinfo{year}{2008}\natexlab{}.
\newblock \showarticletitle{A Scalable, Commodity Data Center Network
  Architecture}. In \bibinfo{booktitle}{\emph{ACM SIGCOMM Computer
  Communication Review}}, Vol.~\bibinfo{volume}{38}. ACM,
  \bibinfo{pages}{63--74}.
\newblock


\bibitem[\protect\citeauthoryear{Alur, Fisman, and Raghothaman}{Alur
  et~al\mbox{.}}{2016}]%
        {alur2016regular}
\bibfield{author}{\bibinfo{person}{Rajeev Alur}, \bibinfo{person}{Dana Fisman},
  {and} \bibinfo{person}{Mukund Raghothaman}.} \bibinfo{year}{2016}\natexlab{}.
\newblock \showarticletitle{Regular programming for quantitative properties of
  data streams}. In \bibinfo{booktitle}{\emph{{ESOP} 2016}}.
  \bibinfo{pages}{15--40}.
\newblock


\bibitem[\protect\citeauthoryear{Anderson, Foster, Guha, Jeannin, Kozen,
  Schlesinger, and Walker}{Anderson et~al\mbox{.}}{2014}]%
        {AFGJKSW13a}
\bibfield{author}{\bibinfo{person}{Carolyn~Jane Anderson},
  \bibinfo{person}{Nate Foster}, \bibinfo{person}{Arjun Guha},
  \bibinfo{person}{Jean-Baptiste Jeannin}, \bibinfo{person}{Dexter Kozen},
  \bibinfo{person}{Cole Schlesinger}, {and} \bibinfo{person}{David Walker}.}
  \bibinfo{year}{2014}\natexlab{}.
\newblock \showarticletitle{{NetKAT}: Semantic Foundations for Networks}. In
  \bibinfo{booktitle}{\emph{POPL}}. \bibinfo{pages}{113--126}.
\newblock


\bibitem[\protect\citeauthoryear{Bhatia, Chen, Boutros, Binderberger, and
  Haas}{Bhatia et~al\mbox{.}}{2014}]%
        {rfc7130}
\bibfield{author}{\bibinfo{person}{Manav Bhatia}, \bibinfo{person}{Mach Chen},
  \bibinfo{person}{Sami Boutros}, \bibinfo{person}{Marc Binderberger}, {and}
  \bibinfo{person}{Jeffrey Haas}.} \bibinfo{year}{2014}\natexlab{}.
\newblock \bibinfo{title}{{Bidirectional Forwarding Detection (BFD) on Link
  Aggregation Group (LAG) Interfaces}}.
\newblock \bibinfo{howpublished}{RFC 7130}.   (\bibinfo{date}{Feb.}
  \bibinfo{year}{2014}).
\newblock
\urldef\tempurl%
\url{https://doi.org/10.17487/RFC7130}
\showDOI{\tempurl}


\bibitem[\protect\citeauthoryear{Davis}{Davis}{2004}]%
        {UMFPACK}
\bibfield{author}{\bibinfo{person}{Timothy~A. Davis}.}
  \bibinfo{year}{2004}\natexlab{}.
\newblock \showarticletitle{Algorithm 832: UMFPACK V4.3---an
  Unsymmetric-pattern Multifrontal Method}.
\newblock \bibinfo{journal}{\emph{ACM Trans. Math. Softw.}}
  \bibinfo{volume}{30}, \bibinfo{number}{2} (\bibinfo{date}{June}
  \bibinfo{year}{2004}), \bibinfo{pages}{196--199}.
\newblock
\urldef\tempurl%
\url{https://doi.org/10.1145/992200.992206}
\showDOI{\tempurl}


\bibitem[\protect\citeauthoryear{Di~Pierro, Hankin, and Wiklicky}{Di~Pierro
  et~al\mbox{.}}{2005}]%
        {di2005probabilistic}
\bibfield{author}{\bibinfo{person}{Alessandra Di~Pierro},
  \bibinfo{person}{Chris Hankin}, {and} \bibinfo{person}{Herbert Wiklicky}.}
  \bibinfo{year}{2005}\natexlab{}.
\newblock \showarticletitle{Probabilistic $\lambda$-calculus and quantitative
  program analysis}.
\newblock \bibinfo{journal}{\emph{Journal of Logic and Computation}}
  \bibinfo{volume}{15}, \bibinfo{number}{2} (\bibinfo{year}{2005}),
  \bibinfo{pages}{159--179}.
\newblock
\urldef\tempurl%
\url{https://doi.org/10.1093/logcom/exi008}
\showDOI{\tempurl}


\bibitem[\protect\citeauthoryear{Droste, Kuich, and Vogler}{Droste
  et~al\mbox{.}}{2009}]%
        {droste2009handbook}
\bibfield{author}{\bibinfo{person}{Manfred Droste}, \bibinfo{person}{Werner
  Kuich}, {and} \bibinfo{person}{Heiko Vogler}.}
  \bibinfo{year}{2009}\natexlab{}.
\newblock \bibinfo{booktitle}{\emph{Handbook of Weighted Automata}}.
\newblock \bibinfo{publisher}{Springer}.
\newblock


\bibitem[\protect\citeauthoryear{Foster, Kozen, Mamouras, Reitblatt, and
  Silva}{Foster et~al\mbox{.}}{2016}]%
        {probnetkat-cantor}
\bibfield{author}{\bibinfo{person}{Nate Foster}, \bibinfo{person}{Dexter
  Kozen}, \bibinfo{person}{Konstantinos Mamouras}, \bibinfo{person}{Mark
  Reitblatt}, {and} \bibinfo{person}{Alexandra Silva}.}
  \bibinfo{year}{2016}\natexlab{}.
\newblock \showarticletitle{Probabilistic {NetKAT}}. In
  \bibinfo{booktitle}{\emph{ESOP}}. \bibinfo{pages}{282--309}.
\newblock
\urldef\tempurl%
\url{https://doi.org/10.1007/978-3-662-49498-1_12}
\showDOI{\tempurl}


\bibitem[\protect\citeauthoryear{Foster, Kozen, Milano, Silva, and
  Thompson}{Foster et~al\mbox{.}}{2015}]%
        {FKMST15a}
\bibfield{author}{\bibinfo{person}{Nate Foster}, \bibinfo{person}{Dexter
  Kozen}, \bibinfo{person}{Matthew Milano}, \bibinfo{person}{Alexandra Silva},
  {and} \bibinfo{person}{Laure Thompson}.} \bibinfo{year}{2015}\natexlab{}.
\newblock \showarticletitle{A Coalgebraic Decision Procedure for {NetKAT}}. In
  \bibinfo{booktitle}{\emph{POPL}}. ACM, \bibinfo{pages}{343--355}.
\newblock


\bibitem[\protect\citeauthoryear{Gill, Jain, and Nagappan}{Gill
  et~al\mbox{.}}{2011}]%
        {gill11}
\bibfield{author}{\bibinfo{person}{Phillipa Gill}, \bibinfo{person}{Navendu
  Jain}, {and} \bibinfo{person}{Nachiappan Nagappan}.}
  \bibinfo{year}{2011}\natexlab{}.
\newblock \showarticletitle{Understanding Network Failures in Data Centers:
  Measurement, Analysis, and Implications}. In \bibinfo{booktitle}{\emph{ACM
  SIGCOMM}}. \bibinfo{pages}{350--361}.
\newblock


\bibitem[\protect\citeauthoryear{Giry}{Giry}{1982}]%
        {giry1982categorical}
\bibfield{author}{\bibinfo{person}{Michele Giry}.}
  \bibinfo{year}{1982}\natexlab{}.
\newblock \showarticletitle{A categorical approach to probability theory}.
\newblock In \bibinfo{booktitle}{\emph{Categorical aspects of topology and
  analysis}}. \bibinfo{publisher}{Springer}, \bibinfo{pages}{68--85}.
\newblock
\urldef\tempurl%
\url{https://doi.org/10.1007/BFb0092872}
\showDOI{\tempurl}


\bibitem[\protect\citeauthoryear{Gordon, Henzinger, Nori, and Rajamani}{Gordon
  et~al\mbox{.}}{2014}]%
        {gordon2014probabilistic}
\bibfield{author}{\bibinfo{person}{Andrew~D Gordon}, \bibinfo{person}{Thomas~A
  Henzinger}, \bibinfo{person}{Aditya~V Nori}, {and} \bibinfo{person}{Sriram~K
  Rajamani}.} \bibinfo{year}{2014}\natexlab{}.
\newblock \showarticletitle{Probabilistic programming}. In
  \bibinfo{booktitle}{\emph{Proceedings of the on Future of Software
  Engineering}}. ACM, \bibinfo{pages}{167--181}.
\newblock
\urldef\tempurl%
\url{https://doi.org/10.1145/2593882.2593900}
\showDOI{\tempurl}


\bibitem[\protect\citeauthoryear{Guo, Lu, Li, Wu, Zhang, Shi, Tian, Zhang, and
  Lu}{Guo et~al\mbox{.}}{2009}]%
        {guo2009bcube}
\bibfield{author}{\bibinfo{person}{Chuanxiong Guo}, \bibinfo{person}{Guohan
  Lu}, \bibinfo{person}{Dan Li}, \bibinfo{person}{Haitao Wu},
  \bibinfo{person}{Xuan Zhang}, \bibinfo{person}{Yunfeng Shi},
  \bibinfo{person}{Chen Tian}, \bibinfo{person}{Yongguang Zhang}, {and}
  \bibinfo{person}{Songwu Lu}.} \bibinfo{year}{2009}\natexlab{}.
\newblock \showarticletitle{BCube: A High Performance, Server-centric Network
  Architecture for Modular Data Centers}.
\newblock \bibinfo{journal}{\emph{ACM SIGCOMM Computer Communication Review}}
  \bibinfo{volume}{39}, \bibinfo{number}{4} (\bibinfo{year}{2009}),
  \bibinfo{pages}{63--74}.
\newblock


\bibitem[\protect\citeauthoryear{Guo, Wu, Tan, Shi, Zhang, and Lu}{Guo
  et~al\mbox{.}}{2008}]%
        {guo2008dcell}
\bibfield{author}{\bibinfo{person}{Chuanxiong Guo}, \bibinfo{person}{Haitao
  Wu}, \bibinfo{person}{Kun Tan}, \bibinfo{person}{Lei Shi},
  \bibinfo{person}{Yongguang Zhang}, {and} \bibinfo{person}{Songwu Lu}.}
  \bibinfo{year}{2008}\natexlab{}.
\newblock \showarticletitle{Dcell: A Scalable and Fault-Tolerant Network
  Structure for Data Centers}. In \bibinfo{booktitle}{\emph{ACM SIGCOMM
  Computer Communication Review}}, Vol.~\bibinfo{volume}{38}. ACM,
  \bibinfo{pages}{75--86}.
\newblock


\bibitem[\protect\citeauthoryear{Kazemian, Varghese, and McKeown}{Kazemian
  et~al\mbox{.}}{2012}]%
        {hsa}
\bibfield{author}{\bibinfo{person}{Peyman Kazemian}, \bibinfo{person}{George
  Varghese}, {and} \bibinfo{person}{Nick McKeown}.}
  \bibinfo{year}{2012}\natexlab{}.
\newblock \showarticletitle{Header Space Analysis: Static Checking for
  Networks}. In \bibinfo{booktitle}{\emph{{USENIX} {NSDI} 2012}}.
  \bibinfo{pages}{113--126}.
\newblock
\showISBNx{978-931971-92-8}
\urldef\tempurl%
\url{https://www.usenix.org/conference/nsdi12/technical-sessions/presentation/kazemian}
\showURL{%
\tempurl}


\bibitem[\protect\citeauthoryear{Kemeny, Snell, et~al\mbox{.}}{Kemeny
  et~al\mbox{.}}{1960}]%
        {kemeny1960finite}
\bibfield{author}{\bibinfo{person}{John~G Kemeny},
  \bibinfo{person}{James~Laurie Snell}, {et~al\mbox{.}}}
  \bibinfo{year}{1960}\natexlab{}.
\newblock \bibinfo{booktitle}{\emph{Finite markov chains}}.
  Vol.~\bibinfo{volume}{356}.
\newblock \bibinfo{publisher}{van Nostrand Princeton, NJ}.
\newblock
\showISBNx{978-0-387-90192-3}


\bibitem[\protect\citeauthoryear{Khurshid, Zhou, Caesar, and Godfrey}{Khurshid
  et~al\mbox{.}}{2012}]%
        {veriflow}
\bibfield{author}{\bibinfo{person}{Ahmed Khurshid}, \bibinfo{person}{Wenxuan
  Zhou}, \bibinfo{person}{Matthew Caesar}, {and} \bibinfo{person}{Brighten
  Godfrey}.} \bibinfo{year}{2012}\natexlab{}.
\newblock \showarticletitle{Veriflow: Verifying Network-Wide Invariants in Real
  Time}. In \bibinfo{booktitle}{\emph{ACM SIGCOMM}}. \bibinfo{pages}{467--472}.
\newblock


\bibitem[\protect\citeauthoryear{Kozen}{Kozen}{1981}]%
        {K81c}
\bibfield{author}{\bibinfo{person}{Dexter Kozen}.}
  \bibinfo{year}{1981}\natexlab{}.
\newblock \showarticletitle{Semantics of probabilistic programs}.
\newblock \bibinfo{journal}{\emph{J. Comput. Syst. Sci.}} \bibinfo{volume}{22},
  \bibinfo{number}{3} (\bibinfo{year}{1981}), \bibinfo{pages}{328--350}.
\newblock
\urldef\tempurl%
\url{https://doi.org/10.1016/0022-0000(81)90036-2}
\showDOI{\tempurl}


\bibitem[\protect\citeauthoryear{Kozen}{Kozen}{1997}]%
        {K97c}
\bibfield{author}{\bibinfo{person}{Dexter Kozen}.}
  \bibinfo{year}{1997}\natexlab{}.
\newblock \showarticletitle{Kleene Algebra with Tests}.
\newblock \bibinfo{journal}{\emph{ACM TOPLAS}} \bibinfo{volume}{19},
  \bibinfo{number}{3} (\bibinfo{date}{May} \bibinfo{year}{1997}),
  \bibinfo{pages}{427--443}.
\newblock
\urldef\tempurl%
\url{https://doi.org/10.1145/256167.256195}
\showDOI{\tempurl}


\bibitem[\protect\citeauthoryear{Kwiatkowska, Norman, and Parker}{Kwiatkowska
  et~al\mbox{.}}{2011}]%
        {kwiatkowska2011prism}
\bibfield{author}{\bibinfo{person}{M. Kwiatkowska}, \bibinfo{person}{G.
  Norman}, {and} \bibinfo{person}{D. Parker}.} \bibinfo{year}{2011}\natexlab{}.
\newblock \showarticletitle{{PRISM} 4.0: Verification of Probabilistic
  Real-time Systems}. In \bibinfo{booktitle}{\emph{Proc. 23rd International
  Conference on Computer Aided Verification (CAV'11)}}
  \emph{(\bibinfo{series}{LNCS})},
  \bibfield{editor}{\bibinfo{person}{G.~Gopalakrishnan} {and}
  \bibinfo{person}{S.~Qadeer}} (Eds.), Vol.~\bibinfo{volume}{6806}.
  \bibinfo{publisher}{Springer}, \bibinfo{pages}{585--591}.
\newblock
\urldef\tempurl%
\url{https://doi.org/10.1007/978-3-642-22110-1_47}
\showDOI{\tempurl}


\bibitem[\protect\citeauthoryear{Liu, Halperin, Krishnamurthy, and
  Anderson}{Liu et~al\mbox{.}}{2013}]%
        {liu2013f10}
\bibfield{author}{\bibinfo{person}{Vincent Liu}, \bibinfo{person}{Daniel
  Halperin}, \bibinfo{person}{Arvind Krishnamurthy}, {and}
  \bibinfo{person}{Thomas~E Anderson}.} \bibinfo{year}{2013}\natexlab{}.
\newblock \showarticletitle{F10: A Fault-Tolerant Engineered Network}. In
  \bibinfo{booktitle}{\emph{{USENIX} {NSDI}}}. \bibinfo{pages}{399--412}.
\newblock


\bibitem[\protect\citeauthoryear{Mai, Khurshid, Agarwal, Caesar, Godfrey, and
  King}{Mai et~al\mbox{.}}{2011}]%
        {anteater}
\bibfield{author}{\bibinfo{person}{Haohui Mai}, \bibinfo{person}{Ahmed
  Khurshid}, \bibinfo{person}{Rachit Agarwal}, \bibinfo{person}{Matthew
  Caesar}, \bibinfo{person}{P.~Brighten Godfrey}, {and}
  \bibinfo{person}{Samuel~Talmadge King}.} \bibinfo{year}{2011}\natexlab{}.
\newblock \showarticletitle{Debugging the Data Plane with {Anteater}}. In
  \bibinfo{booktitle}{\emph{ACM SIGCOMM}}. \bibinfo{pages}{290--301}.
\newblock


\bibitem[\protect\citeauthoryear{Mohri}{Mohri}{2000}]%
        {mohri2000generic}
\bibfield{author}{\bibinfo{person}{Mehryar Mohri}.}
  \bibinfo{year}{2000}\natexlab{}.
\newblock \showarticletitle{Generic $\varepsilon$-removal algorithm for
  weighted automata}. In \bibinfo{booktitle}{\emph{{CIAA} 2000}}. Springer,
  \bibinfo{pages}{230--242}.
\newblock


\bibitem[\protect\citeauthoryear{Niranjan~Mysore, Pamboris, Farrington, Huang,
  Miri, Radhakrishnan, Subramanya, and Vahdat}{Niranjan~Mysore
  et~al\mbox{.}}{2009}]%
        {niranjan2009portland}
\bibfield{author}{\bibinfo{person}{Radhika Niranjan~Mysore},
  \bibinfo{person}{Andreas Pamboris}, \bibinfo{person}{Nathan Farrington},
  \bibinfo{person}{Nelson Huang}, \bibinfo{person}{Pardis Miri},
  \bibinfo{person}{Sivasankar Radhakrishnan}, \bibinfo{person}{Vikram
  Subramanya}, {and} \bibinfo{person}{Amin Vahdat}.}
  \bibinfo{year}{2009}\natexlab{}.
\newblock \showarticletitle{Portland: A Scalable Fault-Tolerant Layer 2 Data
  Center Network Fabric}. In \bibinfo{booktitle}{\emph{ACM SIGCOMM Computer
  Communication Review}}, Vol.~\bibinfo{volume}{39}. ACM,
  \bibinfo{pages}{39--50}.
\newblock


\bibitem[\protect\citeauthoryear{Olejnik, Wiklicky, and Cheraghchi}{Olejnik
  et~al\mbox{.}}{2016}]%
        {olejnik2016probabilistic}
\bibfield{author}{\bibinfo{person}{Maciej Olejnik}, \bibinfo{person}{Herbert
  Wiklicky}, {and} \bibinfo{person}{Mahdi Cheraghchi}.}
  \bibinfo{year}{2016}\natexlab{}.
\newblock \bibinfo{title}{Probabilistic Programming and Discrete Time Markov
  Chains}.
\newblock   (\bibinfo{year}{2016}).
\newblock
\urldef\tempurl%
\url{http://www.imperial.ac.uk/media/imperial-college/faculty-of-engineering/computing/public/MaciejOlejnik.pdf}
\showURL{%
\tempurl}


\bibitem[\protect\citeauthoryear{Roy, Zeng, Bagga, Porter, and Snoeren}{Roy
  et~al\mbox{.}}{2015}]%
        {roy15}
\bibfield{author}{\bibinfo{person}{Arjun Roy}, \bibinfo{person}{Hongyi Zeng},
  \bibinfo{person}{Jasmeet Bagga}, \bibinfo{person}{George Porter}, {and}
  \bibinfo{person}{Alex~C. Snoeren}.} \bibinfo{year}{2015}\natexlab{}.
\newblock \showarticletitle{Inside the Social Network's (Datacenter) Network}.
  In \bibinfo{booktitle}{\emph{ACM SIGCOMM}}. \bibinfo{pages}{123--137}.
\newblock


\bibitem[\protect\citeauthoryear{Saheb-Djahromi}{Saheb-Djahromi}{1980}]%
        {Saheb-Djahromi80}
\bibfield{author}{\bibinfo{person}{N. Saheb-Djahromi}.}
  \bibinfo{year}{1980}\natexlab{}.
\newblock \showarticletitle{{CPO}s of measures for nondeterminism}.
\newblock \bibinfo{journal}{\emph{Theoretical Computer Science}}
  \bibinfo{volume}{12} (\bibinfo{year}{1980}), \bibinfo{pages}{19--37}.
\newblock
\urldef\tempurl%
\url{https://doi.org/10.1016/0304-3975(80)90003-1}
\showDOI{\tempurl}


\bibitem[\protect\citeauthoryear{Sharir, Pnueli, and Hart}{Sharir
  et~al\mbox{.}}{1984}]%
        {sharir1984verification}
\bibfield{author}{\bibinfo{person}{Micha Sharir}, \bibinfo{person}{Amir
  Pnueli}, {and} \bibinfo{person}{Sergiu Hart}.}
  \bibinfo{year}{1984}\natexlab{}.
\newblock \showarticletitle{Verification of probabilistic programs}.
\newblock \bibinfo{journal}{\emph{SIAM J. Comput.}} \bibinfo{volume}{13},
  \bibinfo{number}{2} (\bibinfo{year}{1984}), \bibinfo{pages}{292--314}.
\newblock
\urldef\tempurl%
\url{https://doi.org/10.1137/0213021}
\showDOI{\tempurl}


\bibitem[\protect\citeauthoryear{Singla, Hong, Popa, and Godfrey}{Singla
  et~al\mbox{.}}{2012}]%
        {singla2012jellyfish}
\bibfield{author}{\bibinfo{person}{Ankit Singla}, \bibinfo{person}{Chi-Yao
  Hong}, \bibinfo{person}{Lucian Popa}, {and} \bibinfo{person}{P~Brighten
  Godfrey}.} \bibinfo{year}{2012}\natexlab{}.
\newblock \showarticletitle{Jellyfish: Networking Data Centers Randomly}. In
  \bibinfo{booktitle}{\emph{{USENIX} {NSDI}}}. \bibinfo{pages}{225--238}.
\newblock


\bibitem[\protect\citeauthoryear{Smolka, Eliopoulos, Foster, and Guha}{Smolka
  et~al\mbox{.}}{2015}]%
        {compilekat}
\bibfield{author}{\bibinfo{person}{Steffen Smolka}, \bibinfo{person}{Spiros
  Eliopoulos}, \bibinfo{person}{Nate Foster}, {and} \bibinfo{person}{Arjun
  Guha}.} \bibinfo{year}{2015}\natexlab{}.
\newblock \showarticletitle{A Fast Compiler for {NetKAT}}. In
  \bibinfo{booktitle}{\emph{ICFP 2015}}.
\newblock
\urldef\tempurl%
\url{https://doi.org/10.1145/2784731.2784761}
\showDOI{\tempurl}


\bibitem[\protect\citeauthoryear{Smolka, Kumar, Foster, Kozen, and
  Silva}{Smolka et~al\mbox{.}}{2017}]%
        {probnetkat-scott}
\bibfield{author}{\bibinfo{person}{Steffen Smolka}, \bibinfo{person}{Praveen
  Kumar}, \bibinfo{person}{Nate Foster}, \bibinfo{person}{Dexter Kozen}, {and}
  \bibinfo{person}{Alexandra Silva}.} \bibinfo{year}{2017}\natexlab{}.
\newblock \showarticletitle{{C}antor Meets {S}cott: Semantic Foundations for
  Probabilistic Networks}. In \bibinfo{booktitle}{\emph{{POPL} 2017}}.
\newblock
\urldef\tempurl%
\url{https://doi.org/10.1145/3009837.3009843}
\showDOI{\tempurl}


\bibitem[\protect\citeauthoryear{Tarjan}{Tarjan}{1975}]%
        {Tarjan75}
\bibfield{author}{\bibinfo{person}{Robert~Endre Tarjan}.}
  \bibinfo{year}{1975}\natexlab{}.
\newblock \showarticletitle{Efficiency of a Good But Not Linear Set Union
  Algorithm}.
\newblock \bibinfo{journal}{\emph{J. {ACM}}} \bibinfo{volume}{22},
  \bibinfo{number}{2} (\bibinfo{year}{1975}), \bibinfo{pages}{215--225}.
\newblock
\urldef\tempurl%
\url{https://doi.org/10.1145/321879.321884}
\showDOI{\tempurl}


\bibitem[\protect\citeauthoryear{Valiant}{Valiant}{1982}]%
        {valiant82}
\bibfield{author}{\bibinfo{person}{L. Valiant}.}
  \bibinfo{year}{1982}\natexlab{}.
\newblock \showarticletitle{{A Scheme for Fast Parallel Communication}}.
\newblock \bibinfo{journal}{\emph{SIAM J. Comput.}} \bibinfo{volume}{11},
  \bibinfo{number}{2} (\bibinfo{year}{1982}), \bibinfo{pages}{350--361}.
\newblock


\bibitem[\protect\citeauthoryear{Wang, Hoffmann, and Reps}{Wang
  et~al\mbox{.}}{2018}]%
        {prob-AI}
\bibfield{author}{\bibinfo{person}{Di Wang}, \bibinfo{person}{Jan Hoffmann},
  {and} \bibinfo{person}{Thomas Reps}.} \bibinfo{year}{2018}\natexlab{}.
\newblock \showarticletitle{{PMAF}: An Algebraic Framework for Static Analysis
  of Probabilistic Programs}. In \bibinfo{booktitle}{\emph{{POPL} 2018}}.
\newblock
\urldef\tempurl%
\url{https://www.cs.cmu.edu/~janh/papers/WangHR17.pdf}
\showURL{%
\tempurl}


\end{thebibliography}

\appendix

\section{Omitted Proofs}
\label{app:omitted-proofs}

\begin{lemma}
\label{lem:convergence-on-cantor-open}
Let $A$ be a finite boolean combination of basic open sets, \ie sets of the form
$B_a = \sset{a}\uparrow$ for $a \in \pfin\Hist$, and let $\oldden{-}$ denote
the semantics from \cite{probnetkat-scott}. Then for all programs
$\polp$ and inputs $a \in \pH$,
\begin{align*}
  \oldden{\polp\star}(a)(A) = \lim_{n\to\infty} \oldden{\polp^{(n)}}(a)(A)
\end{align*}
\end{lemma}
\begin{proof*}
Using topological arguments, the claim follows directly from previous results:
$A$ is a Cantor-clopen set by \cite{probnetkat-scott} (\ie, both $A$ and
$\scomp{A}$ are Cantor-open), so its indicator function
$\mathbf{1}_A$ is Cantor-continuous. But $\mu_n \defeq \oldden{\polp^{(n)}}(a)$
converges weakly to $\mu \defeq \oldden{\polp\star}(a)$ in the Cantor topology
(Theorem~4 in \cite{probnetkat-cantor}), so
\begin{align*}
\lim_{n\to\infty} \oldden{\polp^{(n)}}(a)(A)
= \lim_{n\to\infty} \int \mathbf{1}_A d\mu_n
= \int \mathbf{1}_A d\mu
= \oldden{\polp\star}(a)(A)
\end{align*}
(To see why $A$ and $\scomp{A}$ are open in the Cantor topology, note that they
can be written in disjunctive normal form over atoms $B_{\sset{h}}$.)
\end{proof*}

\begin{proof}[Proof of \Cref{prop:old-and-new-sem}]
We only need to show that for $\pdup$-free programs $\polp$ and history-free
inputs $a \in \pPk$, $\oldden{\polp}(a)$ is a distribution on packets (where 
we identify packets and singleton histories).
We proceed by structural induction on $\polp$. All cases are straightforward
except perhaps the case of $\polp\star$. For this case, by the induction hypothesis, all
$\den{\polp^{(n)}}(a)$ are discrete probability distributions on packet sets, therefore vanish outside $\pPk$. By \Cref{lem:convergence-on-cantor-open}, this is also true of the limit $\den{\polp\star}(a)$, as its value on $\pPk$ must be 1, therefore it is also a discrete distribution on packet sets.
\end{proof}

\begin{proof}[Proof of \Cref{lem:ptwise-convergence}]
This follows directly from \Cref{lem:convergence-on-cantor-open} and
\Cref{prop:old-and-new-sem} by noticing that $\emph{any}$ set $A \subseteq \pPk$
is a finite boolean combination of basic open sets.
\end{proof}

\begin{lemma}
\label{lem:inverse}
The matrix $X=I-Q$ in Equation \eqref{eq:inverse} of \S\ref{sec:closed-form} is invertible.
\end{lemma}
\begin{proof*}
Let $S$ be a finite set of states, $\len S=n$, $M$ an $S\times S$ substochastic matrix ($M_{st}\geq 0$, $M\One\leq\One$). A state $s$ is \emph{defective} if $(M\One)_s < 1$. We say $M$ is \emph{stochastic} if $M\One=\One$, \emph{irreducible} if $(\sum_{i=0}^{n-1} M^i)_{st} > 0$ (that is, the support graph of $M$ is strongly connected), and \emph{aperiodic} if all entries of some power of $M$ are strictly positive.

We show that if $M$ is substochastic such that every state can reach a defective state via a path in the support graph, then the spectral radius of $M$ is strictly less than $1$. Intuitively, all weight in the system eventually drains out at the defective states.

Let $e_s$, $s\in S$, be the standard basis vectors. As a distribution, $e_s^T$ is the unit point mass on $s$. For $A\subs S$, let $e_A=\sum_{s\in A} e_s$. The $L_1$-norm of a substochastic vector is its total weight as a distribution. Multiplying on the right by $M$ never increases total weight, but will strictly decrease it if there is nonzero weight on a defective state. Since every state can reach a defective state, this must happen after $n$ steps, thus $\norm{e_s^TM^n}<1$. Let $c = \max_s \norm{e_s^TM^n}<1$.
For any $y=\sum_s a_s e_s$,
\begin{align*}
\norm{y^TM^n} &= \norm{(\sum_s a_s e_s)^TM^n}\\
&\leq \sum_s \len{a_s}\cdot\norm{e_s^TM^n}
\leq \sum_s \len{a_s}\cdot c
= c\cdot\norm{y^T}.
\end{align*}
Then $M^n$ is contractive in the $L_1$ norm, so $\len\lambda < 1$ for all eigenvalues $\lambda$.
Thus $I-M$ is invertible because $1$ is not an eigenvalue of $M$.
\end{proof*}

\section{Encoding 2-Generative Automata in Full \probnetkat}
\label{app:bi-auto}

To keep notation light, we describe our encoding in the special case where the
alphabet $A = \{ x, y \}$, there are four states $S = \{ s_1, s_2, s_3, s_4 \}$,
the initial state is $s_1$, and the output function $\rho$ is
\[
  \rho(s_1) = (x, \blank) \qquad
  \rho(s_2) = (y, \blank) \qquad
  \rho(s_3) = (\blank, x) \qquad
  \rho(s_4) = (\blank, y) .
\]
Encoding general automata is not much more complicated.  Let $\tau : S \to
\Dist(S)$ be a given transition function; we write $p_{i,j}$ for
$\tau(s_i)(s_j)$. We will build a \probnetkat policy simulating this automaton.
Packets have two fields, $\kw{st}$ and $\kw{id}$, where $\kw{st}$ ranges over $S
\cup A \cup \{ \bullet \}$ and $\kw{id}$ ranges over $\{ 1, 2 \}$. Define:
\[
  p \defeq \pseq{\pseq
  {\match{\kw{st}}{s_1}}
  {\pstar{\kw{loop}}}}
  {\modify{\kw{st}}{\bullet}}
\]
The initialization keeps packets that start in the initial state, while the
final command marks histories that have exited the loop by setting $\kw{st}$ to
be special letter $\bullet$.

The main program $\kw{loop}$ first branches on the current state $\kw{st}$:
\[
  \kw{loop} \defeq \text{case } \begin{cases}
    \match{\kw{st}}{s_1} : \kw{state}1 \\
    \match{\kw{st}}{s_2} : \kw{state}2 \\
    \match{\kw{st}}{s_3} : \kw{state}3 \\
    \match{\kw{st}}{s_4} : \kw{state}4
  \end{cases}
\]
Then, the policy simulates the behavior from each state. For instance:
\begin{align*}
  \kw{state1} \defeq \bigoplus \begin{cases}
    \pseq{(
      \ite{\match{\kw{id}}{1}}
      {\pseq {\modify{\kw{st}}{x}} {\pdup}}
      {\ptrue})}
    {\modify{\kw{st}}{s_1}}
    \withp p_{1, 1} , \\
    \pseq{(
      \ite{\match{\kw{id}}{1}}
      {\pseq {\modify{\kw{st}}{y}} {\pdup}}
      {\ptrue})}
    {\modify{\kw{st}}{s_2}}
    \withp p_{1, 2} , \\
    \pseq{(
      \ite{\match{\kw{id}}{2}}
      {\pseq {\modify{\kw{st}}{x}} {\pdup}}
      {\ptrue})}
    {\modify{\kw{st}}{s_3}}
    \withp p_{1, 3} , \\
    \pseq{(
      \ite{\match{\kw{id}}{2}}
      {\pseq {\modify{\kw{st}}{y}} {\pdup}}
      {\ptrue})}
    {\modify{\kw{st}}{s_4}}
    \withp p_{1, 4}
  \end{cases}
\end{align*}
The policies $\kw{state2}, \kw{state3}, \kw{state4}$ are defined similarly.

Now, suppose we are given two 2-generative automata $W, W'$ that differ only in
their transition functions. For simplicity, we will further assume that both
systems have strictly positive probability of generating a letter in either
component in finitely many steps from any state. Suppose they generate
distributions $\mu, \mu'$ respectively over pairs of infinite words $A^\omega
\times A^\omega$. Now, consider the encoded \probnetkat policies $p, p'$. We
argue that $\den{p} = \den{q}$ if and only if $\mu = \mu'$.\footnote{%
  We will not present the semantics of \probnetkat programs with $\pdup$ here;
  instead, the reader should consult earlier
  papers~\cite{probnetkat-cantor,probnetkat-scott} for the full
development.}

First, it can be shown that $\den{p} = \den{p'}$ if and only if $\den{p}(e) =
\den{p'}(e)$, where
\[
  e \defeq \{ \pi \pi \mid \pi \in \Pk \} .
\]
Let $\nu = \den{p}(e)$ and $\nu' = \den{p'}(e)$. The key connection between the
automata and the encoded policies is the following equality:
\begin{equation}
  \label{eq:bi-auto:pnk}
  \mu(S_{u, v}) = \nu(T_{u, v})
\end{equation}
for every pair of finite prefixes $u, v \in A^*$. In the automata distribution
on the left, $S_{u, v} \subseteq A^\omega \times A^\omega$ consists of all pairs
of infinite strings where $u$ is a prefix of the first component and $v$ is a
prefix of the second component. In the \probnetkat distribution on the right,
we first encode $u$ and $v$ as packet histories. For $i \in \{ 1, 2 \}$
representing the component and $w \in A^*$ a finite word, define the history
\[
  \h_i(w) \in \Hist \defeq
    (\kw{st} = \bullet, \kw{id} = i)
    , (\kw{st} = w[|w|], \kw{id} = i) 
    , \dots
    , (\kw{st} = w[1], \kw{id} = i) 
    , (\kw{st} = s_1, \kw{id} = i) .
\]
The letters of the word $w$ are encoded in reverse order because by convention,
the head/newest packet is written towards the left-most end of a packet history,
while the oldest packet is written towards the right-most end. For instance, the
final letter $w[|w|]$ is the most recent (\ie, the latest) letter produced by
the policy. Then, $T_{u, v}$ is the set of all history sets including $\h_1(u)$
and $\h_2(v)$:
\[
  T_{u, v} \defeq \{ a \in \pH \mid \h_1(u) \in a , \h_2(v) \in a \} .
\]
Now $\den{p} = \den{p'}$ implies $\mu = \mu'$, since \Cref{eq:bi-auto:pnk} gives
\[
  \mu(S_{u, v}) = \mu'(S_{u, v}) .
\]
The reverse implication is a bit more delicate.  Again by \Cref{eq:bi-auto:pnk},
we have
\[
  \nu(T_{u, v}) = \nu'(T_{u, v}) .
\]
We need to extend this equality to all cones, defined by packet histories $\h$:
\[
  B_{\h} \defeq \{ a \in \pH \mid \h \in a \} .
\]
This follows by expressing $B_{\h}$ as boolean combinations of $T_{u, v}$, and
observing that the encoded policy produces only sets of encoded histories, \ie,
where the most recent state $\kw{st}$ is set to $\bullet$ and the initial state
$\kw{st}$ is set to $s_1$.


\end{document}